\documentclass[12pt,preprintnumbers,nofootinbib,letterpaper]{article}
\pdfoutput=1
\usepackage{amsmath,amssymb,amsfonts,cite,color,epsf,epsfig,epstopdf,graphics,graphicx,mathtools}
\usepackage{cite}
\definecolor{ultramarine}{rgb}{0.07, 0.04, 0.56}
\definecolor{cadmiumgreen}{rgb}{0.0, 0.42, 0.24}
\definecolor{indigo(dye)}{rgb}{0.0, 0.25, 0.42}
\usepackage[linktocpage=true,breaklinks]{hyperref}
\usepackage[utf8]{inputenc}

\hypersetup{
colorlinks=true,
citecolor=ultramarine,
linkcolor=cadmiumgreen,
urlcolor=indigo(dye),
}

\begin{document}

\begin{flushright} {\footnotesize YITP-20-118, IPMU20-0099}  \end{flushright}
\vspace{0.5cm}
\begin{center}

{\Large\bf Stealth dark energy in scordatura DHOST theory}
\\[1.2cm]

{Mohammad Ali Gorji$^{1}$, 
	Hayato Motohashi$^{2}$,
	Shinji Mukohyama$^{1,3}$}
\\[.7cm]

{\small \textit{$^1$Center for Gravitational Physics, Yukawa Institute for Theoretical
		Physics, Kyoto University, \\ Kyoto 606-8502, Japan
}}\\

{\small \textit{$^2$Division of Liberal Arts, Kogakuin University,
2665-1 Nakano-machi, Hachioji, \\ Tokyo 192-0015, Japan
}}\\

{\small \textit{$^3$Kavli Institute for the Physics and Mathematics of the Universe (WPI), The University of Tokyo, \\ Chiba 277-8583, Japan}}\\

\end{center}
\vspace{.8cm}
\hrule \vspace{0.3cm}

\begin{abstract}
A stealth de Sitter solution in scalar-tensor theories has an exact de Sitter background metric and a nontrivial scalar field profile.  Recently, in the context of Degenerate Higher-Order Scalar-Tensor (DHOST) theories it was shown that stealth de Sitter solutions suffer from either infinite strong coupling or gradient instability for scalar field perturbations. The sound speed squared is either vanishing or negative. In the first case, the strong coupling scale is zero and thus lower than the energy scale of any physical phenomena. From the viewpoint of effective field theory, this issue is naturally resolved by introducing a controlled detuning of the degeneracy condition dubbed scordatura, recovering a version of ghost condensation. In this paper we construct a viable dark energy model in the scordatura DHOST theory based on a stealth cosmological solution, in which the metric is the same as in the standard $\Lambda$CDM model and the scalar field profile is linearly time-dependent.  We show that the scordatura mechanism resolves the strong coupling and gradient instability.  Further, we find that the scordatura is also necessary to make the quasi-static limit well-defined, which implies that the subhorizon observables are inevitably affected by the scordatura. We derive the effective gravitational coupling and the correction to the friction term for the subhorizon evolution of the linear dark matter energy density contrast as well as the Weyl potential and the gravitational slip parameter. In the absence of the scordatura, the quasi-static approximation would break down at all scales around stealth cosmological solutions even if the issue of the infinite strong coupling is unjustly disregarded. Therefore previous estimations of the subhorizon evolution of matter density contrast in modified gravity in the literature need to be revisited by taking into account the scordatura effect. 
\end{abstract}
\vspace{0.5cm} \hrule
\setcounter{footnote}{0}
\newpage

\section{Introduction}

It is well known that taking the effects of extra dimensions and/or higher-derivative curvature terms into account, the gravitational action would generally be endowed with scalar degree(s) of freedom in addition to the two tensor degrees of freedom. This scalar degree of freedom, corresponding to the longitudinal mode of gravity, can be realized by considering a scalar field in the gravitational action leading to scalar-tensor theories. 
In general, including higher derivatives would produce the so-called Ostrogradsky ghost which leads to an unbounded Hamiltonian \cite{Ostrogradsky:1850fid,Woodard:2015zca,Motohashi:2014opa,Motohashi:2020psc,Aoki:2020gfv}.
The Ostrogradsky ghost can be avoided by requiring the Lagrangian to satisfy a certain set of degeneracy conditions \cite{Langlois:2015cwa,Motohashi:2016ftl,Motohashi:2017eya,Motohashi:2018pxg}. 
Degenerate Higher-Order Scalar-Tensor (DHOST) theories \cite{Langlois:2015cwa,Crisostomi:2016czh,BenAchour:2016fzp} provide a general framework, including the previously known higher-derivative scalar-tensor theories such as Horndeski \cite{Horndeski:1974wa} and beyond Horndeski theories \cite{Zumalacarregui:2013pma,Gleyzes:2014dya}, which were constructed by systematically taking possible higher-derivative interactions into account and requiring the degeneracy condition to evade the Ostrogradsky ghost.  Further generalization is possible by requiring the absence of the Ostrogradsky ghost only in the unitary gauge~\cite{Gao:2014soa,DeFelice:2018mkq,Gao:2018znj,Motohashi:2020wxj}. Apart from the systematic construction based on the degeneracy condition, the higher-derivative terms can be simply taken into account by performing disformal transformations on a scalar-tensor theory without higher derivatives \cite{Zumalacarregui:2012us,Bettoni:2013diz}. It is also shown that DHOST theories can be obtained by performing disformal transformations on the Horndeski and beyond Horndeski theories \cite{Zumalacarregui:2013pma,Gleyzes:2014dya,Crisostomi:2016czh,Achour:2016rkg,BenAchour:2016fzp} as invertible transformations keep the number of propagating degrees of freedom unchanged~\cite{Domenech:2015tca,Takahashi:2017zgr}. Moreover, looking at the singular limit of the disformal transformations \cite{Deruelle:2014zza}, new class of scalar-tensor theories known as mimetic theories \cite{Chamseddine:2013kea} appear which do not belong to the standard setup of the DHOST theories \cite{Takahashi:2017pje,Langlois:2018jdg}. However, the mimetic scalar-tensor theories usually suffer from the gradient/ghost instabilities \cite{Firouzjahi:2017txv,Zheng:2017qfs,Hirano:2017zox,Gorji:2017cai,Takahashi:2017pje} and generally suffer from the caustic singularities \cite{Mukohyama:2009tp,Barvinsky:2013mea,Gorji:2019rlm}.

The effects of the scalar field in scalar-tensor theories can be modeled as an effective energy-momentum tensor which sources the standard Einstein-Hilbert action. The stealth solutions are those scalar field background configurations such that the background effective energy-momentum tensor takes the form of that for the cosmological constant. In this case, the background metric takes the same form as that in General Relativity (GR) as an exact solution of the background equations of motion. It was recently shown in \cite{Motohashi:2018wdq,Takahashi:2020hso} that, if a scalar-tensor theory satisfies a certain set of conditions, it is possible to accommodate any metric solution same as in GR with general matter component as an exact solution of the equations of motion while the whole effects of the scalar degree of freedom at the background level are simply to shift the cosmological constant with respect to the value of the bare cosmological constant originating from the zero point energy. For instance, the ghost condensation setup provides a stealth Minkowski solution \cite{ArkaniHamed:2003uy} and stealth Schwarzschild solution \cite{Mukohyama:2005rw}.  For a stealth de Sitter solution in DHOST theories see \cite{Crisostomi:2018bsp}, and for stealth black hole solutions in scalar-tensor theories see Refs.~\cite{Babichev:2013cya,Minamitsuji:2018vuw,BenAchour:2018dap,Motohashi:2019sen,Charmousis:2019vnf,Minamitsuji:2019shy,BenAchour:2019fdf,Minamitsuji:2019tet,Bernardo:2019yxp,Minamitsuji:2019tet,Takahashi:2019oxz,Gorji:2019rlm,Takahashi:2020hso,BenAchour:2020fgy}, among which the most general case, the Kerr-Newman-de Sitter solution in the DHOST theories, was obtained in \cite{Takahashi:2020hso}. The difference between the standard GR solutions and stealth solutions in scalar-tensor theories arises at the level of perturbations due to the scalar field perturbations. In this regard, one can find stealth background solutions whose geometries are the same as well-known cosmological and black hole background solutions in GR and then look for the effects of modified gravity by studying perturbations. 

However, in the context of DHOST theories it turned out that the scalar mode around the stealth solutions is in general strongly coupled \cite{Minamitsuji:2018vuw,deRham:2019slh,Motohashi:2019ymr,Khoury:2020aya}. In particular, it was shown in \cite{Motohashi:2019ymr} that either strong coupling or gradient instability is inevitable for asymptotically flat or de Sitter stealth solutions so long as the evolution equation of linear perturbations is second order. Since the DHOST theories are designed to satisfy the degeneracy condition at the nonlinear level to prevent the Ostrogradsky ghost \cite{Langlois:2015cwa,Langlois:2015skt}, the dispersion relation for the scalar field perturbations takes the standard linear form $\omega = {\bar c}_s k$, where ${\bar c}_s$ is the sound speed, and the Ostrogradsky ghost does not propagate accordingly \cite{Langlois:2017mxy}. Around the stealth solution, however, the sound speed squared ${\bar c}_s^2$ is actually either negative or vanishing. For the latter case, the strong coupling scale was shown to be much lower than the energy scale of the dynamics of the scalar field perturbation \cite{Motohashi:2019ymr}. The same logic holds for a broader framework of the effective field theory (EFT) so long as the evolution equation of perturbations is second order.  Hence, the scalar perturbations around asymptotically flat or de Sitter stealth solutions universally suffer from either gradient instability or infinite strong coupling.  Such solutions cannot be trusted as they are beyond the regime of validity of the EFT of ghost-free higher-derivative theory.

This issue can be circumvented by the scordatura mechanism~\cite{Motohashi:2019ymr}, namely, by introducing a controlled detuning of the degeneracy condition from the EFT point of view.  It renders the perturbations weakly coupled with the cost of a benign apparent Ostrogradsky ghost, which propagates only above the EFT cutoff scale. If we do not impose the degeneracy condition, the higher-order terms provide nonlinear dispersion relation of the form $\omega^2 \ni k^{2n}$ with $n>1$ ($n=2$ for ghost condensation~\cite{ArkaniHamed:2003uy}, $n=3$ for Ho\v{r}ava-Lisfshitz gravity~\cite{Horava:2009uw} and other $n$ as well~\cite{Motohashi:2019ymr,Langlois:2017mxy,Motohashi:2020wxj}). Since the strong coupling scale is determined by a positive power of the sound speed, one would expect that this modification helps to resolve the issue of strong coupling. However, if the degeneracy condition is largely violated, the Ostrogradsky ghost would propagate even below the EFT cutoff and make the setup unstable. On the other hand, there is no fundamental symmetry associated to the degeneracy condition and it will be finally broken by quantum corrections. Therefore, small deviation from the degeneracy condition is expected to be inevitable from the quantum point of view. The realization of this fact at the level of an EFT leads to a version of the ghost condensation \cite{ArkaniHamed:2003uy}, dubbed the scordatura theory \cite{Motohashi:2019ymr}. In this scenario, first, it was shown that the dispersion relation for the scalar field perturbations around the stealth solution takes the nonlinear form $(\omega/M) \approx \sqrt{\alpha}\,{\cal O}(1) (k/aM)^2$. Here, $a(t)$ is the scale factor, $M$ is the EFT cutoff scale determined by the vacuum expectation value of the scalar field, and $\alpha$ is a dimensionless parameter, which determines the deviation from the degeneracy condition so that for $\alpha=0$ the setup reduces to the DHOST theories.  Using the EFT framework, the strong coupling scale was shown to be $\sim |\alpha|^{7/2}M$ in the presence of the scordatura term.  Hence, by choosing $\alpha={\cal O}(1)$, the strong coupling scale is as high as the EFT cutoff scale~\cite{ArkaniHamed:2003uy}.  Second, it was shown that for a small deviation from the degeneracy condition, the mass of the Ostrogradsky ghost mode is larger than the EFT cutoff scale $M$. Thus, the scordatura scenario allows us to find an EFT where the scalar field perturbations are weakly coupled and there is no Ostrogradsky ghost all the way up to the scale $M$.

Since the scordatura of the degeneracy condition inevitably appears from the EFT point of view, it is important to clarify its effects on observables.  Having weakly-coupled stealth solutions in the scordatura scenario, we can look for their phenomenological applications in cosmology and black hole physics.  For stealth black hole solution, so long as the deviation from the degeneracy condition is under control, the evolution of the stealth black hole is expected to be sufficiently slow.  Indeed, it was shown in \cite{Mukohyama:2005rw} that in the case of ghost condensate the accretion of the scalar field for the stealth black hole is sufficiently slow so that it cannot be distinguished by observations. Nevertheless, the scordatura is important for the recovery of the generalized second law of black hole thermodynamics as well~\cite{Mukohyama:2009rk,Mukohyama:2009um}. On the other hand, it is nontrivial how the scordatura changes the evolution of perturbations around cosmological stealth solutions (see \cite{Mukohyama:2006be} for a preliminary study of this question). Thus, one interesting possibility is to construct a dark energy model using a background stealth solution same as in the $\Lambda$CDM model and to investigate phenomenology at the level of perturbations, which is the main topic of the present paper.

The rest of the paper is organized as follows. In Section \ref{sec-scordatura} we present our particular scordatura theory as a small deviation from a subset of DHOST theories. In Sections \ref{sec-cos-BG} and \ref{sec-Cos-pert} we study cosmological background and perturbations of the model, respectively. In Section \ref{sec-stealth-DE} we construct our stealth dark energy model and show that it is weakly-coupled by virtue of the scordatura term. In Section \ref{sec-coupling-DM}, we study cosmological perturbations of the system of the dark energy minimally coupled to the dark matter and find the effective gravitational coupling for the matter energy density contrast as well as corrections to the friction term and the gravitational slip parameter. Section \ref{sec-summary} is devoted to the summary and conclusions. Finally, we present some technical details in appendices \ref{appendix} and \ref{appendix-b}.

\section{Scordatura theory: Small deviation from DHOST}\label{sec-scordatura}

In this section, we present the action for the scordatura model which is investigated in \cite{Motohashi:2019ymr}. Let us start with an action with quadratic derivative interactions
\begin{eqnarray}\label{action-DHOST}
S_{\rm D} = \int d^4 x \sqrt{-g} \Big[ F_0(\phi,X) + F_1(\phi,X) \Box \phi + F_2(\phi,X) R 
+ \sum_{i=1}^5 A_i(\phi,X) L_i \Big] \,,
\end{eqnarray}
where $X=g^{\nu\eta}\phi_{\nu}\phi_{\eta}$ with $\phi_{\nu}\equiv\nabla_{\nu}\phi$, $F_0(\phi,X)$ is the k-essence term~\cite{ArmendarizPicon:1999rj}, $F_1(\phi,X)$ corresponds to the so-called kinetic-braiding term~\cite{Deffayet:2010qz}, and $F_2(\phi,X)$ is the non-minimal kinetic coupling of the scalar field to the curvature. The functions $A_i(\phi,X)$ label the quadratic higher-derivative Lagrangian densities which are defined as
\begin{align}\label{DHOST-L2s}
L_1 &= \phi_{\nu\eta} \phi^{\nu\eta} , \hspace{1cm} L_2 = (\Box \phi)^2 ,
\hspace{1cm} L_3 = \Box\phi \, \phi_{\nu}\phi^{\nu\eta} \phi_{\eta} , 
\nonumber \\
L_4 &= \phi^{\nu} \phi_{\nu\eta} \phi^{\eta\lambda}\phi_{\lambda} , 
\hspace{1cm} L_5 = (\phi_{\nu}\phi^{\nu\eta}\phi_{\eta})^2 \,.
\end{align}

For general functional forms of functions $F_i$ and $A_i$, the Ostrogradsky ghost propagates in the model \eqref{action-DHOST}. However, requiring functions $F_i$ and $A_i$ to satisfy a set of degeneracy conditions in the context of DHOST theories, the model possesses only three degrees of freedom and becomes free from the Ostrogradsky ghost \cite{Langlois:2015cwa}. 

Assuming the shift-symmetry for the action (\ref{action-DHOST}) with $F_i(X)$, $A_i(X)$, and a linearly time-dependent scalar field in a cosmological setup, the model supports an exact de Sitter solution \cite{Crisostomi:2018bsp}. The reason is simple since for that scalar field profile, $X$ becomes constant and all the second and higher time derivatives of scalar field vanish. Therefore, we have an exact de Sitter solution. This solution, however, is stealth in the sense that all the effects of the scalar field at the background level is to shift the cosmological constant. In other words, the scalar field does not gravitate at the background level while it has non-trivial stress-energy tensor at the level of perturbations. 
Note that it was shown in \cite{Minamitsuji:2018vuw} that in general the shift symmetry is not essential and it is possible to accommodate stealth solutions in a class of shift symmetry breaking theories.
Moreover, recently, various stealth solution in DHOST theories have been found: the Schwarzschild-(anti-)de Sitter solution \cite{BenAchour:2018dap,Motohashi:2019sen,Minamitsuji:2019shy}, general static spherically symmetric solution \cite{Takahashi:2019oxz}, Kerr-de Sitter solution \cite{Charmousis:2019vnf}, and finally it was shown that any solution same as in GR in the presence of general matter component can be accommodated in a class of DHOST theories \cite{Takahashi:2020hso}, generalizing the result for a constant scalar field profile \cite{Motohashi:2018wdq}.
Although the scalar field perturbations around the stealth solution are free of pathological ghosts, they are actually infinitely strongly coupled as we explained in the introduction \cite{Minamitsuji:2018vuw,deRham:2019slh,Motohashi:2019ymr,Khoury:2020aya}. 
In particular, using the EFT framework it was shown in \cite{Motohashi:2019ymr} that either strong coupling or gradient instability is inevitable for asymptotically de Sitter or flat stealth solutions so long as the evolution equation of perturbations is second order.

This issue can be remedied in the scordatura scenario~\cite{Motohashi:2019ymr}, where a controlled detuning of the degeneracy condition is introduced from the EFT point of view. 
A deviation from the degeneracy condition can be achieved by adding any of the quadratic higher-derivative Lagrangian densities in \eqref{DHOST-L2s} with free coefficients to the DHOST Lagrangian. Among them, $L_1$ and $L_2$ give dominant contributions to the dispersion relation of perturbation around the stealth de Sitter solution. We can see this by looking at the numbers of time and spatial derivatives in $L_i$ which yield $\omega$ and $k$ contribution to the dispersion relation, respectively. For the linear perturbations corresponding to the quadratic action, we find that $L_1$ and $L_2$ provide $\omega^4, \omega^2k^2$ and $k^4$ terms, $L_3$ provides $\omega^4, \omega^2k^2$, $H \omega^3$ and $H\omega k^2$ terms, $L_4$ gives $\omega^4$ and $\omega^2k^2$ terms, and $L_5$ can only give $\omega^4$ term. Higher-derivative terms are suppressed by the Planck mass (see appendix~\ref{appendix}), and hence only leading-order contribution is important. Around the stealth solution, the dispersion relation takes the form of $\omega^2\sim \bar c_s^2k^2$ with $\bar c_s^2\ll 1$, so long as the EOM for perturbation is second order~\cite{Motohashi:2019ymr}, which applies to the DHOST theories satisfying the degeneracy condition at the nonlinear level. Hence, the leading-order correction is $k^4$ originating from $L_1$ and $L_2$, rendering the dispersion relation of the form $\omega^2 \sim k^4/M^2$, where $M$ is the cutoff scale of the EFT. (With this dispersion relation, other terms in $L_1$ and $L_2$ are subdominant as $\omega^4\sim k^8/M^4 \ll k^4$ and $\omega^2k^2\sim k^6/M^2\ll k^4$ below the cutoff.) Moreover, the Lagrangian density $L_1$ contributes to the deviation of the speed of gravitational waves from the speed of light, which is severely constrained~\cite{Langlois:2017dyl} by the recent gravitational waves observations~\cite{Monitor:2017mdv}. Therefore, we choose $L_2$ to slightly break the degeneracy condition and consider the following action
\begin{eqnarray}\label{action-scordatura}
S_{\rm g} = \int d^4x \sqrt{-g}L_{\rm g} 
= \int d^4 x \sqrt{-g} \bigg[ F_0(X) + F_1(X) \Box\phi + F_2(X) R
+ \frac{6 F_{2,X}^2}{F_2} \phi^{\nu} \phi_{\nu\eta} \phi^{\eta\lambda}\phi_{\lambda}
- \frac{\alpha}{2} \frac{(\Box\phi)^2}{M^2} \bigg] \,,
\end{eqnarray}
where $F_{2,X}=\partial F_2/\partial X$, and we have assumed the shift symmetry for simplicity. Here, $M$ is a mass scale corresponding to the cutoff scale of the EFT, and $\alpha$ is a dimensionless parameter governing the scordatura term.  The above action is a shift-symmetric subset of action~(\ref{action-DHOST}) with $A_2 = -\frac{\alpha}{2 M^2}$, $A_4 = \frac{6 F_{2,X}^2}{F_2}$, and $A_i = 0$ for $i=1,3,5$. For $\alpha=0$, the degeneracy conditions of the quadratic DHOST theories~\cite{Langlois:2015cwa} are satisfied while for $\alpha\neq0$ the scordatura term slightly breaks the degeneracy condition to avoid the strong coupling or the gradient instability. Of course, such a violation of the degeneracy condition summons the Ostrogradsky ghost but it shows up only above the EFT cutoff scale, as far as $\alpha$ is non-vanishing and of order unity. Related to the DHOST part of the action~\eqref{action-scordatura}, as we mentioned above, the term $A_1$ contributes to the speed of gravitational waves, which is severely constrained by the recent gravitational waves observations~\cite{Langlois:2017dyl,Monitor:2017mdv} and we then set $A_1=0$. We also set $A_3=0$, which implies $A_5=0$ from the degeneracy condition, to prevent the graviton decay~\cite{Creminelli:2018xsv}.

Taking variation of the action~\eqref{action-scordatura} with respect to the metric, we find the Einstein equations
\begin{equation}\label{EEs}
M_P^2 G_{\nu\eta} = T_{\nu\eta} \, \hspace{1cm} \mbox{with} \hspace{1cm}
T_{\nu\eta} \equiv -\frac{2}{\sqrt{-g}} \frac{\delta (\sqrt{-g}L_{\phi} ) }{\delta g^{\nu\eta}} + g_{\nu\eta} L_{\phi} \,,
\end{equation}
where $M_P=(8\pi{G})^{-1/2}$ is the reduced Planck mass\footnote{We work in the unit with $c=1=\hbar$, where $c$ is the speed of light in vacuum and $\hbar$ is the reduced Planck constant.}, $G_{\nu\eta}$ is the Einstein's tensor, $T_{\nu\eta}$ is the effective energy-momentum tensor for the scalar field. Here $L_{\phi} \equiv L_{\rm g} - \frac{M_P^2}{2} R$ is defined so that the total gravitational Lagrangian density $L_{\rm g}$ defined by (\ref{action-scordatura}) reduces to the standard Einstein-Hilbert action in the absence of the scalar field $\phi$ and for $F_2=M_P^2/2$. The explicit form of the effective energy-momentum tensor is complicated and we do not present it here.

Variation with respect to the scalar field also gives the equation of motion
\begin{equation}\label{KG-abstract}
\frac{1}{\sqrt{-g}}\partial_{\nu} (\sqrt{-g}J^{\nu}) = 0 \,, \hspace{1cm} \mbox{with} \hspace{1cm} J^{\nu} \equiv \frac{\delta S_{\rm g}}{\delta\phi_{\nu}} = \frac{\partial L_{\rm g}}{\partial \phi_{\nu}} - \nabla_{\eta} \Big(\frac{\partial L_{\rm g}}{\partial \phi_{\nu\eta}}\Big) \,,
\end{equation}
where $J^\nu$ is the conserved Noether current associated to the shift symmetry whose explicit form is again complicated which we do not present here.

\section{Cosmological background}\label{sec-cos-BG}

Having obtained the equations of motion for the metric and the scalar field, in this section, we find cosmological background equations for the model. For this purpose, in appendix~\ref{appendix}, considering the symmetries of the cosmological background, we systematically defined the relevant dimensionless coordinates, which are given by
\begin{equation}\label{coordinets-ch}
{\tilde t} \equiv \mu M t \,, \hspace{1cm} {\tilde x}^i \equiv \mu M x^i\,; \hspace{1cm} 
\mu\equiv {M/M_P}\,,
\end{equation}
and also the following dimensionless quantities
\begin{equation}\label{dimensionless-couplings}
\phi \equiv M_P \, \varphi\,, \hspace{.5cm} X\equiv{M^4}{\mathrm x}\,, \hspace{.5cm} 
F_0 \equiv M^4 f_0 \,, \hspace{.5cm}
F_1 \equiv M f_1 \,, \hspace{.5cm} F_2 \equiv M_P^2 f_2 \,.
\end{equation}
Here we assume that $\mu\ll 1$, namely, the EFT cutoff scale is sufficiently lower than the Planck scale.

The spatially flat Friedmann-Lema\^{i}tre-Robertson-Walker (FLRW) background in terms of the dimensionless coordinates \eqref{coordinets-ch} takes the form
\begin{equation}\label{metric-FRW-BG}
ds^2 = \frac{M_P^{2}}{M^4} \Big[ - d{\tilde t}^2 + a({\tilde t})^2 \delta_{ij} d{\tilde x}^i d{\tilde x}^j \Big] \,,
\end{equation}
where $a({\tilde t})$ is the scale factor, ${\tilde t}$ and ${\tilde x}^i$ with $i=1,2,3$ are the dimensionless cosmic time and dimensionless spatial coordinates, respectively. 

The dimensionless scalar field $\varphi$ defined in \eqref{dimensionless-couplings} also acquires a homogeneous time-dependent vacuum expectation value $\langle\varphi\rangle=\varphi({\tilde t})$ in the cosmological background Eq.~\eqref{metric-FRW-BG}.

In terms of these new dimensionless quantities, for the background geometry (\ref{metric-FRW-BG}), the temporal component of the Einstein equations (\ref{EEs}) gives the first Friedmann equation
\begin{eqnarray}\label{Friedmann}
&&f_0 + 2 \dot{\varphi}^2 f_{0,{\mathrm x}} + 6 f_2 h^2
+ 12 \dot{\varphi}^2 f_{2,{\mathrm x}} (2 h^2 + \dot{h} )
- 12 \dot{\varphi} \ddot{\varphi} f_{2,{\mathrm x}} h
- 6 \mu \dot{\varphi}^3 f_{1,{\mathrm x}} h
 \\ \nonumber
&&- 6 \dot{\varphi}^2 \frac{ f_{2,{\mathrm x}}^2}{f_2} 
\Big( \ddot{\varphi}^2 + 2 \dddot{\varphi} \dot{\varphi} + 6 h \dot{\varphi} \ddot{\varphi}
+ 2  (\dot{\varphi}\ddot{\varphi})^2 \Big( \frac{f_{2,{\mathrm x}}}{f_2} 
- 2 \frac{f_{2,{\mathrm x}{\mathrm x}}}{f_{2,{\mathrm x}}} \Big)
\Big)
+ \frac{\alpha}{2} \mu^2 \Big( (\Box\varphi)^2+2\dot{\varphi}(\Box\varphi)\dot{} \Big) 
= 0 \,,
\end{eqnarray}
where a dot denotes a derivative with respect to the dimensionless cosmic time ${\tilde t}$, $h\equiv \dot{a}/a$ is the dimensionless Hubble parameter which is related to the standard Hubble parameter as $H = \mu {M} h$, and $\Box\varphi$ is the d'Alembertian defined in the spirit of dimensionless cosmic time ${\tilde t}$ given by
\begin{equation}\label{Boxphi-a}
\Box\varphi = - \frac{1}{a^3} \big(a^3 \dot{\varphi}\big)\dot{} \,.
\end{equation}
On the other hand, the spatial components give the second Friedmann equation
\begin{eqnarray}\label{Raychuadhuri}
&&f_0 + 2f_2 (2\dot{h}+3h^2) - 4 f_{2,{\mathrm x}} \Big( (\dot{\varphi}\ddot{\varphi})\dot{} 
+ h \dot{\varphi} \ddot{\varphi} + 2 \dot{\varphi}^2 \ddot{\varphi}^2 
\Big(\frac{3f_{2,{\mathrm x}}}{4f_2} - \frac{f_{2,{\mathrm x}{\mathrm x}}}{f_{2,{\mathrm x}}} \Big) \Big) 
\nonumber \\ 
&& - 2 \mu f_{1,{\mathrm x}} \dot{\varphi}^2 \ddot{\varphi} 
+ \frac{\alpha}{2} \mu^2 \Big( (\Box\varphi)^2-2\dot{\varphi}(\Box\varphi)\dot{} \Big) 
= 0 \,.
\end{eqnarray}
While we did not explicitly show the functionality of scalar functions in (\ref{Friedmann}) and (\ref{Raychuadhuri}), we should keep in mind that they are functions of time only at the level of the background.

The equation of motion for the scalar field~(\ref{KG-abstract}) then yields
\begin{equation}\label{KG}
\frac{1}{a^3} \big( a^3 j^0 \big)\dot{} = 0 \,,
\end{equation}
where
\begin{align}\label{j0}
j^0 &= -2 f_{0,{\mathrm x}} \dot{\varphi} - 12 f_{2,{\mathrm x}} \dot{\varphi} (2h^2 + \dot{h}) 
+ 6 \mu f_{1,{\mathrm x}} \dot{\varphi}^2 h 
+ \alpha \mu^2 (\Box\varphi)\dot{} \\ \nonumber
&~~~ + 12 \dot{\varphi} \frac{f_{2,{\mathrm x}}^2}{f_2} \Big( 
(\dot{\varphi}\ddot{\varphi})\dot{} + 3h \dot{\varphi}\ddot{\varphi}
+ (\dot{\varphi}\ddot{\varphi})^2 \Big( \frac{f_{2,{\mathrm x}}}{f_2} - 2 \frac{f_{2,{\mathrm x}{\mathrm x}}}{f_{2,{\mathrm x}}} \Big)
\Big) \,,
\end{align}
is the temporal component of the dimensionless current defined as $j^\mu \equiv J^\mu/M^2$ with $J^\mu$ defined in \eqref{KG-abstract}. For the canonical scalar field with $f_0 = -{\mathrm x}/2$, $f_2 =1/2$, and $f_1=0=\alpha$, from the above relation we find $j^0=\dot{\varphi}$. Then Eq.~(\ref{KG}) reduces to the well-known Klein-Gordon equation $\Box\varphi = 0$ for the canonical scalar field.

We see that the third time derivative of $\varphi$ together with the second time derivative of the scale factor encoded in $\dot{h}$ appear in the Friedmann equations (\ref{Friedmann}) and (\ref{Raychuadhuri}) even in the absence of the scordatura term $\alpha=0$. However, by taking a linear combination of these equations we can obtain a second-order equation for $\varphi$, which only includes $h$ but not $\dot{h}$. This means that there is no Ostrogradsky ghost for the DHOST subset with $\alpha=0$ at the background level. In order to see this fact explicitly, we perform the following transformation at the background level \cite{Crisostomi:2018bsp}
\begin{equation}\label{b}
 b \equiv \sqrt{f_2} \, a \,,
\end{equation}
and, moreover, we define the following dimensionless quantities \cite{Bellini:2014fua,Gleyzes:2014rba,Langlois:2017mxy} (with correction found in Appendix A of \cite{Motohashi:2017gqb})
\begin{equation}\label{alpha-i}
\alpha_H \equiv - {\rm x} \frac{f_{2,{\rm x}}}{f_2}\,, \hspace{1cm}
\alpha_B \equiv \frac{\mu}{2} \frac{\dot{\varphi}\,{\rm x}}{h_b} \frac{f_{1,{\rm x}}}{f_2} + \alpha_H \,, \hspace{1cm}
\alpha_K \equiv - \frac{{\rm x}}{6h_b^2}\frac{f_{0,{\rm x}}}{f_2} + \alpha_H + \alpha_B \,,
\end{equation}
which are first order in derivative of the functions $f_i$ with respect to ${\rm x}$.  As we will see below, these types of dimensionless quantities significantly simplify not only the background equations but also the perturbation analysis in the next section.

The first Friedmann equation~(\ref{Friedmann}) in terms of the new scale factor (\ref{b}) and the dimensionless quantities \eqref{alpha-i} takes the following simple form
\begin{eqnarray}\label{Friedmann-b}
&&f_0 + 6 f_2 \Big( h_b^2 (1 + 2\alpha_K) 
+ 2 \Big( \dot{h}_b + h_b \frac{\ddot{\varphi}}{\dot{\varphi}} \alpha_B \Big) \alpha_H \Big)
 + \frac{\alpha}{2} \mu^2 \Big( (\Box\varphi)^2+2\dot{\varphi}(\Box\varphi)\dot{} \Big) 
= 0 \,,
\end{eqnarray}
where 
\begin{equation}\label{Hubble-b}
h_b \equiv \frac{\dot{b}}{b} 
= h - \frac{\ddot{\varphi}}{\dot{\varphi}} \alpha_H \,,
\end{equation}
is the dimensionless Hubble parameter defined with respect to the new scale factor $b$. The explicit expression for $\Box\varphi$ can be obtained by substituting (\ref{b}) and (\ref{alpha-i}) in (\ref{Boxphi-a}). 

Substituting (\ref{b}) and (\ref{alpha-i}), the second Friedmann equation~\eqref{Raychuadhuri} also takes the simple form
\begin{eqnarray}\label{Raychuadhuri-b}
f_0 + 2f_2 \Big( 3 h_b^2 + 2 \Big( \dot{h}_b + h_b \frac{\ddot{\varphi}}{\dot{\varphi}} \alpha_B \Big) \Big)
+ \frac{\alpha}{2} \mu^2 \Big( (\Box\varphi)^2-2\dot{\varphi}(\Box\varphi)\dot{} \Big) 
= 0 \,.
\end{eqnarray}

Performing the transformation (\ref{b}) and substituting \eqref{alpha-i}, it is straightforward to show that (\ref{j0}) takes the following simple form
\begin{equation}\label{j0-b-0}
j^0 = 12 f_2 \frac{\dot{\varphi}}{{\rm x}} \left( \alpha_H \Big( \dot{h}_b + h_b \frac{\ddot{\varphi}}{\dot{\varphi}} \alpha_B \Big) + h_b^2 \alpha_K \right)
+ \alpha \mu^2 (\Box\varphi)\dot{} \,.
\end{equation}

We note that the effects from the scordatura term in background equations.~(\ref{Friedmann-b}), (\ref{Raychuadhuri-b}), and \eqref{j0-b-0} are suppressed by the Planck scale through the factor $\alpha\mu^2$ and, therefore, we can neglect them at the leading order. 
Thus, in the rest of this section, we set $\alpha=0$ and focus on the DHOST subset of the background equations without the scordatura contribution.
However, the existence of the scordatura corrections is essential to solve the issue of the strong coupling or the gradient instability at the level of perturbations~\cite{Motohashi:2019ymr} as we will see in the next section.  

Neglecting Planck-suppressed scordatura corrections in the background equations by setting $\alpha=0$, we only need to look at the DHOST subset. In this case, we see that in the Friedmann equations~(\ref{Friedmann-b}) and (\ref{Raychuadhuri-b}), $\dot{h}_b$ and $\ddot{\varphi}$ appear only via a special combination $\dot{h}_b + h_b \frac{\ddot{\varphi}}{\dot{\varphi}} \alpha_B$. By making use of this fact, we can take a linear combination of the Friedmann equations (\ref{Friedmann-b}) and (\ref{Raychuadhuri-b}) with $\alpha=0$ to find
\begin{align}
\label{Eq-red} &f_0 + 6 f_2 h_b^2 \bigg(1+\frac{2\alpha_K}{1-3\alpha_H}\bigg) = 0 \,, \\
\label{hb-dot} &\dot{h}_b = \frac{3 h_b^2 \alpha_K}{1-3\alpha_H} - h_b \frac{\ddot \varphi}{\dot \varphi} \alpha_B \,.
\end{align}
The first equation~\eqref{Eq-red} only includes the first derivative of the scalar field $\dot{\varphi}$ and the Hubble parameter $h_b$. Taking a time derivative of Eq.~\eqref{Eq-red} and erasing $\dot{h}_b$ by plugging the second equation~\eqref{hb-dot},
we find a second-order equation of motion for the scalar field. This is a particular realization of the degeneracy in the DHOST Lagrangian in the case of homogeneous background equations. However, this second-order equation for the scalar field has a complicated form and we instead work with the equivalent compact equation~(\ref{KG}) with $j^0$ given in Eq.~\eqref{j0-b-0}. Substituting Eq.~\eqref{hb-dot} into \eqref{j0-b-0} to erase $\dot{h}_b$ and $\ddot{\varphi}$, we find the following simple expression for the shift-symmetry current
\begin{eqnarray}\label{j0-b}
j^0 = 12 f_2 h_b^2 \frac{\dot{\varphi} }{{\rm x}} \frac{\alpha_K}{1-3\alpha_H} \,.
\end{eqnarray}

Now, substituting the above expression in Eq.~\eqref{KG}, we find $\ddot{\varphi}=\ddot{\varphi}(\dot{\varphi},h_b,\dot{h}_b)$, which after substituting Eq.~\eqref{hb-dot} gives a second-order equation of motion for the scalar field $\ddot{\varphi}=\ddot{\varphi}(\dot{\varphi},h_b)$. This is similar to the Klein-Gordon equation for the canonical scalar field but here with more parameters. Taking a time derivative of the result $\ddot{\varphi}=\ddot{\varphi}(\dot{\varphi},h_b)$ we obtain an expression for the third-order time derivative as $\dddot{\varphi}=\dddot{\varphi}(\dot{\varphi},\ddot{\varphi},h_b,\dot{h}_b)$.
Then, substituting Eq.~\eqref{hb-dot} and $\ddot{\varphi}=\ddot{\varphi}(\dot{\varphi},h_b)$, we can express the third derivative of the scalar field purely in terms of the first derivative of the scalar field and the Hubble parameter as $\dddot{\varphi}=\dddot{\varphi}(\dot{\varphi},h_b)$. We do not write down the explicit form of this relation here as it is complicated. We shall use this equation to simplify the action for the perturbations in the next section.

\section{Cosmological perturbations}\label{sec-Cos-pert}

In this section, we study linear scalar cosmological perturbations around the background geometry~(\ref{metric-FRW-BG}). As we have shown in the appendix~\ref{appendix}, in terms of the dimensionless variables, in the comoving gauge where we turn off the scalar field perturbations, the line element for the scalar perturbations is given by
\begin{equation}\label{metric-FRW-Perturbations}
ds^2 = \frac{M_P^2}{M^4} \Big( - ( 1 + 2 A ) d{\tilde t}^2 
+ 2 {\tilde \partial}_i B d{\tilde t} d{\tilde x}^i 
+ a^2 ( 1 + 2 \psi ) \delta_{ij} d{\tilde x}^i d{\tilde x}^j \Big) \,,
\end{equation}
where $(A,B,\psi)$ are scalar perturbations which depend on the dimensionless coordinates $({\tilde t},{\tilde x}^i)$ and ${\tilde \partial}_i$ is a derivative with respect to ${\tilde x}^i$. 

Before providing detailed analysis of the scalar perturbations in Section~\ref{ssec-dho} and \ref{ssec-sco}, here we provide the outline of our analysis.
Substituting (\ref{metric-FRW-Perturbations}) in (\ref{action-scordatura}), expanding the action up to the quadratic order in perturbations, and performing some integration by parts, we obtain the quadratic action for the scalar perturbations as
\begin{equation}\label{action-SS-bare}
S^{(2)}_{\rm g} \equiv \int d{\tilde t} d^3{\tilde x} M^4 {\tilde {\cal L}}^{(2)}_{\rm g} 
\equiv \int d{\tilde t} d^3{\tilde x} M^4 
\Big[ {\tilde {\cal L}}_{\rm D}^{(2)}(\dot{\psi},\psi,\dot{A},A,B) 
+ \alpha \mu^2 {\tilde {\cal L}}_{\rm S}^{(2)}(\dot{\psi},\psi,\dot{A},A,B) \Big] \,,
\end{equation}
where we have defined dimensionless quadratic gravitational Lagrangian ${\tilde {\cal L}}^{(2)}_{\rm g}(\dot{\psi},\psi,\dot{A},A,B)$ and then we decomposed it to the DHOST and scordatura parts as follows 
\begin{equation}\label{Lagrangians-def}
{\tilde {\cal L}}_{\rm D}^{(2)} \equiv {\tilde {\cal L}}^{(2)}_{\rm g}|_{\alpha=0} \,, 
\hspace{1cm} \mbox{and} \hspace{1cm}
{\tilde {\cal L}}_{\rm S}^{(2)} \equiv
\frac{1}{\alpha\mu^2}\Big( {\tilde {\cal L}}^{(2)}_{\rm g} - {\tilde {\cal L}}_{\rm D}^{(2)} \Big) \,.
\end{equation}

From the explicit forms of the quadratic action (\ref{action-SS-bare}), we see that we cannot remove $\dot{A}$ by integration by parts even in the DHOST part and one may naively think that there is an extra degree of freedom even in the DHOST part. However, we already knew that this is not the case as the DHOST theories provide only three degrees of freedom at the nonlinear level \cite{Langlois:2015skt}. To see this fact explicitly in our particular case, we consider the following field redefinition
\begin{equation}\label{xi-def}
\zeta \equiv \psi + \alpha_H A \,,
\end{equation}
which is the counterpart of the background transformation (\ref{b}) for scale factor $b$ but now at the level of perturbations. Performing the above field redefinition, the quadratic action~(\ref{action-SS-bare}) becomes
\begin{equation}\label{action-SS-bare-xi}
S^{(2)}_{\rm g} = \int d{\tilde t} d^3{\tilde x} M^4 
\Big[ {\tilde {\cal L}}_{\rm D}^{(2)}(\dot{\zeta},\zeta,A,B) 
+ \alpha \mu^2 {\tilde {\cal L}}_{\rm S}^{(2)}(\dot{\zeta},\zeta,\dot{A},A,B ) \Big] \,.
\end{equation}
From the above action we see that the DHOST part~${\tilde {\cal L}}_{\rm D}^{(2)}$ provides one dynamical degree of freedom $\zeta$ in the scalar perturbation sector.  We shall provide a detailed analysis of the DHOST part in Section~\ref{ssec-dho}.
On the other hand, the scordatura part~${\tilde {\cal L}}_{\rm S}^{(2)}$ apparently provides one extra dynamical degree of freedom $A$ on top of $\zeta$, which is nothing but the (would-be) Ostrogradsky ghost. It is shown in \cite{Motohashi:2019ymr} that for small deviations from the degeneracy conditions, this (would-be) ghost mode is massive and is excited only for energy scales larger than $M$. Therefore, the setup is free of this type of pathology as far as we work in the energy scales below the EFT cutoff $M$. Thus, the apparent ghost is benign. This result provides a simple way to integrate out not only non-dynamical field $B$ but also apparently dynamical field $A$ as we will show in Section~\ref{ssec-sco}.
After examining each part, in Section~\ref{ssec-gra}, we shall come back to the total quadratic Lagrangian to see the dispersion relation of the scalar perturbation.

\subsection{DHOST part}
\label{ssec-dho}

In this subsection, we present detailed analysis of the DHOST part~${\tilde {\cal L}}_{\rm D}^{(2)}$ of the quadratic action~\eqref{action-SS-bare-xi} for the scalar perturbations, which is necessary for the analysis of perturbations in the scordatura part as we will see in the next subsection.

Substituting (\ref{metric-FRW-Perturbations}) in the action (\ref{action-scordatura}) and then using the dimensionless quantities defined in (\ref{dimensionless-couplings}), it is straightforward to show that the DHOST part~${\tilde {\cal L}}_{\rm D}^{(2)}$ of the quadratic action~\eqref{action-SS-bare-xi} takes the following form
\begin{eqnarray}\label{Lagrangian-DHOST}
{\tilde {\cal L}}_{\rm D}^{(2)} &=& 2 f_2 \Big(
- 3 a^3 \dot{\zeta}^2 + 6 a^3 h_b (1+ \alpha_B) \dot{\zeta} A
- 2 a {\tilde k}^2 \dot{\zeta} B
+ a {\tilde k}^2 \zeta^2 \\ \nonumber
&& \hspace{1cm} + 2 a (1+ \alpha_H) {\tilde k}^2\zeta A 
- 3 a^3 h_b^2 \beta_K A^2 + 2 a h_b (1+\alpha_B) {\tilde k}^2 AB \Big) \,,
\end{eqnarray}
where we have defined dimensionless parameters 
\begin{eqnarray}\label{beta-K}
&&\beta_K \equiv - \frac{{\rm x}^2}{3} \frac{f_{0,{\rm x}{\rm x}} }{h_b^2 f_2} 
+ (1-\alpha_H) (1+3 \alpha_B) + \beta_B 
+ \frac{(1 + 6 \alpha_H - 3 \alpha_H^2) \alpha_K
- 2 ( 2 - 6 \alpha_H + 3 \alpha_K ) \beta_H}{1-3 \alpha_H} \,,
\nonumber \\
&& \beta_B \equiv \mu \dot{\varphi}\,{\rm x}^2 \frac{f_{1,{\rm x}{\rm x}}}{h_b f_2} \,,
\hspace{1cm}
\beta_H \equiv {\rm x}^2 \frac{f_{2,{\rm x}{\rm x}}}{ f_2}
\,,
\end{eqnarray}
which are second order in derivative of the functions $f_i$ with respect to ${\rm x}$. In obtaining the Lagrangian (\ref{Lagrangian-DHOST}), we have used background equations to express $\dot{h}_b$, $\ddot{\varphi}$, and $\dddot{\varphi}$ in terms of $\dot{\varphi}$ and $h_b$ through the way that we explained at the end of Section~\ref{sec-cos-BG}.

As we explained above, the DHOST part quadratic Lagrangian~\eqref{Lagrangian-DHOST} does not contain derivative of $A$ and $B$ and hence they are non-dynamical fields. 
Varying the Lagrangian (\ref{Lagrangian-DHOST}) with respect to $A$ and $B$, we obtain
\begin{align}
A &= \frac{1}{1+\alpha_B} \frac{\dot{\zeta}}{h_b} \,, \label{A-sol}\\
B &= - 3 \bigg[ 1 - \frac{\beta_K}{(1+\alpha_B)^2} \bigg] \frac{a^2}{{\tilde k}^2} \, \dot{\zeta}
- \frac{1+\alpha_H}{1+\alpha_B} \, \frac{\zeta}{h_b} \,.\label{Phi-B-sol}
\end{align}
Substituting the above results into the DHOST Lagrangian~\eqref{Lagrangian-DHOST} and performing an integration by parts, we find the reduced quadratic Lagrangian
\begin{equation}\label{Lagrangian-DHOST-red}
{\tilde {\cal L}}_{\rm D}^{(2)}
= a^3 f_2 \bigg( \bar{\cal A} \, \dot{\zeta}^2 - \bar{\cal B} \, \frac{{\tilde k}^2}{a^2} \zeta^2
\bigg) \,,
\end{equation}
where we have defined
\begin{eqnarray}\label{coefficients-AD-BD}
\bar{\cal A} = 6 \bigg[ 1 - \frac{\beta_K}{(1+\alpha_B)^2} \bigg] \,, \hspace{1cm}
 \bar{\cal B} = - 2 \bigg[ 1 - \frac{1}{a f_2} \frac{d}{d{\tilde t}} 
\bigg(\frac{af_2}{h_b}\frac{1+\alpha_H}{1+\alpha_B}\bigg) \bigg] \,.
\end{eqnarray}

To obtain the dispersion relation, we consider $\zeta\propto e^{-i \omega t}$ in the DHOST quadratic Lagrangian~\eqref{Lagrangian-DHOST-red} and find
\begin{equation}\label{DR-DHOST}
\omega = \bar{c}_s \Big(\frac{k}{a}\Big) \,; \hspace{2cm} 
\bar{c}_s^2 \equiv \frac{\bar{\cal B}}{\bar{\cal A}} \,,
\end{equation}
where $\bar{c}_s$ is the sound speed and $k = \mu M {\tilde k}$ is the dimensionful momentum. We see that although there are higher-derivative terms in the DHOST action (\ref{action-DHOST}) and the equation of motion is apparently higher order, the degeneracy condition led to the standard linear dispersion relation  (\ref{DR-DHOST}).

Taking variation of (\ref{Lagrangian-DHOST-red}), we find the equation of motion for $\zeta$ as
\begin{equation}\label{EoM-zeta}
\ddot{\zeta} + \Big( 3 h + \frac{d}{d{\tilde t}} \ln(f_2 \bar{\cal A} ) \Big) \dot{\zeta} 
+ \Big( \frac{\bar{c}_s {\tilde k}}{a}\Big)^2 \zeta = 0 \,.
\end{equation}

One can use the setup to study dark energy when the de Sitter solution arises thanks to the shift symmetry of the setup. The setup, however, becomes strongly-coupled for the stealth solutions as we will explicitly show in Section~\ref{sec-stealth-DE}. Taking the effects of the scordatura term into account, which makes the dispersion relation nonlinear, we will show that the perturbation becomes weakly-coupled \cite{Motohashi:2019ymr}.

\subsection{Scordatura part}
\label{ssec-sco}

Next, we study the scordatura part~${\tilde {\cal L}}_{\rm S}^{(2)}$ of the quadratic action~\eqref{action-SS-bare-xi}.
Substituting (\ref{metric-FRW-Perturbations}) in the action (\ref{action-scordatura}) and then expanding it up to the quadratic order in scalar modes, it is straightforward to show that the scordatura quadratic Lagrangian is given by
\begin{eqnarray}\label{Lagrangian-scordatura}
{\tilde {\cal L}}_{\rm S}^{(2)} &=& \frac{1}{2} \Big(
{\bar k}_{11} \dot{\zeta}^2 + {\bar k}_{22} \dot{A}^2 + 2 {\bar k}_{12} \dot{\zeta} \dot{A}
+ 2 \dot{\zeta} ({\bar n}_{12} A + {\bar n}_{13} {\tilde k}^2 B) + 2 {\bar n}_{23} {\tilde k}^2 \dot{A} B
\nonumber \\
&& \hspace{.7cm} - {\bar m}_{11} \zeta^2 - 2 {\bar m}_{12} \zeta A - {\bar m}_{22} A^2 
- 2 {\bar m}_{23} {\tilde k}^2 A B + {\bar m}_{33} {\tilde k}^2 B^2 + {\bar m}_{33{\rm s}} {\tilde k}^4 B^2
 \Big) \,,
\end{eqnarray}
where we have defined the following coefficients
\begin{eqnarray}\label{matrices-components-scordatura}
&& {\bar k}_{11} = 9 a^3 {\rm x} , \hspace{2.5cm}
{\bar k}_{12} = - 3 a^3 {\rm x} (1+3\alpha_H) , \hspace{2.5cm} 
{\bar k}_{22} = a^3 {\rm x} (1+3\alpha_H)^2 , \\[15pt] \nonumber
&& {\bar n}_{12} = - 6 a^3 \big( 3{\rm x} h_b 
- (1+6\alpha_H + 3 \alpha_{H}^2-3\beta_H) \dot{\varphi} \ddot{\varphi} \big) , \hspace{.5cm}
{\bar n}_{13} = 3 a {\rm x} , \hspace{.5cm} 
{\bar n}_{23} = - a {\rm x} (1+3\alpha_H) , \\[10pt] \nonumber
&& {\bar m}_{11}  = - \frac{3}{2} a^3 \Big( - 3 {\rm x} ( 2\dot{h}_b  + 3 h_b^2 ) 
+ 6 ( 2 + 3 \alpha_H ) h_b \dot{\varphi} \ddot{\varphi}
\\ \nonumber 
&& \hspace{2.2cm} + ( 1 + 18 \alpha_H + 21 \alpha_H^2 
- 12 \beta_H ) \ddot{\varphi}^2 
+ 2 ( 1 + 3 \alpha_H ) \dot{\varphi} \dddot{\varphi} \Big) , 
\\ \nonumber
&& {\bar m}_{12} = - \frac{3}{2} a^3 \Big( 
- 9 ( 1 - \alpha_H ) h_b^2 {\rm x} + 6 ( 1 + \alpha_H ) {\rm x} \dot{h}_b
 + 6 (1-3\alpha_H) \alpha_H h_b \dot{\varphi} \ddot{\varphi}
\\ \nonumber
&& \hspace{2.25cm} + \big( 1 - \big( 7 + 21 \alpha_H + 21 \alpha_H^2 
- 12 \beta_H \big) \alpha_H + 12 \beta_H \big) \ddot{\varphi}^2
- 2 ( 1 + \alpha_H ) ( 1 + 3 \alpha_H ) \dot{\varphi}  \dddot{\varphi}
\Big) , \\ \nonumber
&& {\bar m}_{22} = 
- \frac{1}{2} a^3 \Big( 9 \big( 5 - 6 \alpha_H - 3 \alpha_H^2 \big) {\rm x} h_b^2 
- 6 \big( 5 + 12 \alpha_H + 3 \alpha_H^2 \big) {\rm x} \dot{h}_b 
\\ \nonumber
&& 
\hspace{2.3cm} - 18 h_b \big( ( 1 - 12 \alpha_H - 9 \alpha_H^2
+ 6 \beta_H ) \alpha_H - 2 \beta_H \big) \dot{\varphi} \ddot{\varphi} 
\\ \nonumber
&& \hspace{2.3cm} + ( 3 \gamma_H - 5) \ddot{\varphi}^2  
+ 2 ( 1 + 3 \alpha_H ) 
\big( 5 + 18 \alpha_H + 9 \alpha_H^2 
- 6 \beta_H \big) \dot{\varphi} \dddot{\varphi} \Big) ,
\\ \nonumber
&& {\bar m}_{23} = - 2 a \Big( - 3 h_b {\rm x} 
+ \big( 1 + 6 \alpha_H + 3 \alpha_H^2 - 3 \beta_H \big) \dot{\varphi} \ddot{\varphi}  \Big) \,, \hspace{1cm}
{\bar m}_{33{\rm s}} = \frac{{\rm x}}{a} \,, \nonumber \\ \nonumber
&& {\bar m}_{33} = \frac{1}{2} a \Big( - 9 h_b^2 {\rm x} + 6 {\rm x} \dot{h}_b 
+ 18 h_b \alpha_H \dot{\varphi} \ddot{\varphi} 
+ \big( 1 - 6 \alpha_H -3 \alpha_H^2 
+ 12 \beta_H \big) \ddot{\varphi}^2 - 2 ( 1 + 3 \alpha_H ) \dot{\varphi} \dddot{\varphi} \Big) \,,
\end{eqnarray}
and also a new dimensionless parameter
\begin{eqnarray}\label{gamma-H}
\gamma_H \equiv - 8 {\rm x}^3 ( 1 + 3 \alpha_H ) \frac{f_{2,{\rm x}{\rm x}{\rm x}}}{f_2}
+ 105 \alpha_H^4 + 208 \alpha_H^3 + 10 \alpha_H^2 ( 11 - 12 \beta_H ) 
+ 16 \alpha_H ( 1 - 9 \beta_H ) - 40 \beta_H ,
\end{eqnarray}
which is third order in derivative of the functions $f_i$ with respect to ${\rm x}$. In obtaining the Lagrangian~(\ref{Lagrangian-scordatura}) we did not use the background equations.

From the scordatura quadratic Lagrangian~(\ref{Lagrangian-scordatura}), we see that it is not possible to remove the time derivative of the field $A$ by integration by parts, which signals the existence of the apparent Ostrogradsky ghost. However, as we already explained, this apparent ghost mode is heavy and propagates only above the EFT cutoff scale $M$ while we are interested in energy scales below $M$ where the only propagating mode is $\zeta$~\cite{Motohashi:2019ymr}. We therefore implement the following strategy to integrate out this heavy mode. Making use of the hierarchy $\mu\ll 1$, we first restrict ourselves to the DHOST quadratic Lagrangian, where we have already found solutions $A=A(\dot{\zeta},\zeta)$ in \eqref{A-sol} and $B=B(\dot{\zeta},\zeta)$ in \eqref{Phi-B-sol}. Since we are not interested in the evolution of the apparent Ostrogradsky ghost showing up only above the cutoff scale $M$, we can substitute the DHOST solutions $A=A(\dot{\zeta},\zeta)$ and $B=B(\dot{\zeta},\zeta)$ into the scordatura action ${\tilde {\cal L}}_{\rm S}^{(2)}$ in \eqref{Lagrangian-scordatura}. 
Moreover, we can erase $\dot A$ in \eqref{Lagrangian-scordatura} as follows.  First we take a time derivative of $A$ in \eqref{A-sol}, and then substituting \eqref{EoM-zeta}, we obtain the following expression up to the first order of $\alpha$
\begin{eqnarray}\label{Phidot}
\dot{A} = - \frac{1}{h_b(1+\alpha_B)} \bigg(
\frac{d}{d{\tilde t}} \Big[ \ln \big(h_b (1+\alpha_B) a^3 f_2 \bar{\cal A} \big)\Big] \, \dot{\zeta}
+ \Big( \frac{\bar{c}_s {\tilde k}}{a}\Big)^2 \zeta 
\bigg) \,.
\end{eqnarray}
Substituting \eqref{A-sol}, \eqref{Phi-B-sol} and \eqref{Phidot} into (\ref{Lagrangian-scordatura}), we can erase both scalar modes $A$ and $B$ in the scordatura quadratic Lagrangian. This method is very efficient while we should always keep in mind that we are completely ignoring the effects of the scordatura term in the equations of motion of $A$ and $B$. This would be consistent in the sense that these corrections are always Planck-suppressed, if we were interested in the non-dynamical variables $A$ and $B$ only. On the contrary, what we are interested in is the reduced action after eliminating $A$ and $B$ in favor of dynamical variables. Actually, as we will show below, we may find artificial infrared divergences in the reduced action due to neglecting these suppressed terms in the equations for $A$ and $B$. Therefore, the consistent way is to keep those scordatura corrections that affect the final result and to neglect only those terms that do not affect the final result. For the computation of the reduced action, this turns out to be equivalent to the following simplified strategy: we first neglect all of the scordatura effects in the equations of motion of $A$ and $B$, and then look at the reduced action to check whether there is an infrared divergence or not in each term of the action; for those terms where there are no infrared divergences, we do not need to do anything more and the leading-order calculation is consistent; for those terms where we find any infrared divergences, on the other hand, the leading-order calculation is not valid and we need to be more careful to take into account the scordatura contributions by looking for term(s) in the scordatura part that cure the artificial divergences. 

Let us now implement the method for our particular case. We see that substituting (\ref{Phi-B-sol}) in (\ref{Lagrangian-scordatura}) to erase $B$, the term that includes ${\bar m}_{33}$ provides ${\tilde {\cal L}}_{\rm S}^{(2)} \supset {\tilde k}^{-2}\dot{\zeta}^2$ which is divergent in the infrared limit ${\tilde k}\to0$. However, the solutions \eqref{A-sol} and (\ref{Phi-B-sol}) are obtained in the absence of the scordatura corrections $\alpha=0$ and, in the presence of the scordatura term, the Lagrangian (\ref{Lagrangian-scordatura}) changes the equations of motion of $A$ and $B$. Looking at the action \eqref{Lagrangian-scordatura}, we see that taking into account the effects of the ${\bar m}_{33}$ in obtaining the equations of motion of the non-dynamical mode $B$ is enough to solve this apparent problem. We therefore do not need to take into account the effects of other terms in the scordatura action \eqref{Lagrangian-scordatura} in the equation of motion of $B$, and also any contribution from the scordatura term in the equation of motion of $A$. Doing so, it is easy to find the correction to the solution of the mode $B$ coming from the scordatura ${\bar m}_{33}$ term as follows
\begin{equation}\label{B-sol}
B = - 3 \bigg[ 1 - \frac{\beta_K}{(1+\alpha_B)^2} \bigg] \frac{a^2\dot{\zeta}}{
{\tilde k}^2+\alpha k_{\rm IR}^2}
- \frac{1+\alpha_H}{1+\alpha_B} \, \frac{\zeta}{h_b} \,; \hspace{1cm} 
k_{\rm IR}^2 \equiv \frac{3\mu^2 a}{4f_2} \frac{\beta_K{\bar m}_{33}}{(1+\alpha_B)^2} \,,
\end{equation}
where the explicit form of ${\bar m}_{33}$ is given by \eqref{matrices-components-scordatura}. 
Comparing \eqref{B-sol} with \eqref{Phi-B-sol}, we see that the denominator ${\tilde k}^2$ is now replaced to ${\tilde k}^2+\alpha k_{\rm IR}^2$ by virtue of the scordatura contribution.
Assuming $\alpha={\cal O}(1)$, for the modes ${\tilde k}\gg k_{\rm IR}$, we see that \eqref{B-sol} recovers the previous result~\eqref{Phi-B-sol}, whereas for the modes ${\tilde k}\ll k_{\rm IR}$, the apparent infrared divergence disappears.
Plugging \eqref{B-sol} into the scordatura Lagrangian~\eqref{Lagrangian-scordatura}, we thus obtain the correct expression without the apparent infrared divergence.
Hence, it is clear that the apparent infrared divergence originates from neglecting a scordatura contribution, which becomes actually leading order at the infrared regime.

Once more, in obtaining the solution \eqref{B-sol}, we only kept the correction from the ${\bar m}_{33}$ term in \eqref{Lagrangian-scordatura} that we need to solve the apparent infrared divergence. We have neglected all other terms since the contributions from them remain subdominant. Of course, we can keep all other terms instead of neglecting them and perform similar calculations.  However, in this case the calculation is much more involved, and at the leading order of $\alpha \mu^2$ the outcome coincides with the one we obtained above.  For instance, if we keep all the terms, the solution~\eqref{A-sol} for $A$ would acquire a scordatura correction, after which, however, we would neglect the scordatura correction since it remains subleading.  We then end up with \eqref{A-sol}.  The strategy we employ here is to include only necessary scordatura correction we should not neglect.  It is an economical way to solve the apparent problem while keeping calculations as simple as possible.

Now, substituting \eqref{A-sol}, \eqref{Phidot}, and \eqref{B-sol} into (\ref{Lagrangian-scordatura}), we find the quadratic scordatura Lagrangian in terms of the single scalar mode $\zeta$ as follows
\begin{equation}\label{Lagrangian-scordatura-red}
{\tilde {\cal L}}_{\rm S}^{(2)} = \frac{a^3}{2} \bigg[ \Big( {\cal A}_1 
+ \frac{a^2 {\cal A}_{2}}{{\tilde k}^2+\alpha {k}_{\rm IR}^2} \Big) \, \dot{\zeta}^2 
- \Big( {\cal B}_{1} \Big(\frac{{\tilde k}}{a}\Big)^2 + {\cal B}_{2}
\Big(\frac{{\tilde k}}{a}\Big)^4 + {\cal M}\Big) \,\zeta^2
\bigg] \,,
\end{equation}
where we have defined five new coefficients ${\cal A}_{1}, {\cal A}_{2}, {\cal B}_{1}, {\cal B}_{2}$, and ${\cal M}$. It is straightforward to find explicit expressions of these coefficients in terms of the dimensionless quantities $\alpha_i$, $\beta_i$ and $\gamma_H$. However, they take complicated forms and hence we do not write their explicit forms here. The scordatura term is expected to play an important role only around the stealth solution where these coefficients take simple forms. Therefore, we only write the explicit forms of ${\cal A}_{1}, {\cal A}_{2}, {\cal B}_{1}, {\cal B}_{2}$, and ${\cal M}$ around the stealth solution in the next section, which is sufficient for our purpose in the present paper.

\subsection{Quadratic gravitational action}
\label{ssec-gra}

Now we combine the results we obtained in Sections~\ref{ssec-dho} and \ref{ssec-sco}.
The total quadratic gravitational Lagrangian~\eqref{action-SS-bare-xi} is given by the sum of the DHOST Lagrangian~(\ref{Lagrangian-DHOST-red}) and the scordatura Lagrangian~(\ref{Lagrangian-scordatura-red}), which turns out to be
\begin{equation}\label{Lagrangian-red}
{\tilde {\cal L}}^{(2)}_{\rm g}
= a^3 f_2 \, {\cal K} \bigg[ \dot{\zeta}^2 
- \Big( c_{\rm s}^2 ({\tilde k})\frac{{\tilde k}^2}{a^2} + \alpha m^2 \Big) \zeta^2
\bigg] \,,
\end{equation}
where we have defined the effective scale-dependent sound speed square (in the sense of a phase velocity) as
\begin{eqnarray}\label{cs2}
c_{\rm s}^2 ({\tilde k}) \equiv \bar{c}_s^2 
+ \frac{\alpha\mu^2}{2f_2} \bigg(
\frac{{\cal B}_{1}}{\bar{\cal A}} - \bar{c}_s^2 \frac{{\cal A}_{1}}{\bar{\cal A}}
+ \Big(\frac{{\tilde k}}{a}\Big)^{2} \frac{{\cal B}_{2}}{\bar{{\cal A}}}
\bigg) \,,
\end{eqnarray}
the mass term
\begin{eqnarray}\label{mass}
m^2 \equiv \frac{\mu^2}{2f_2}
\bigg(
\frac{{\cal M}}{\bar{\cal A}} 
- \bar{c}_s^2 \frac{{\cal A}_{2}}{\bar{\cal A}}
\bigg) \,,
\end{eqnarray}
and the kinetic term coefficient
\begin{equation}\label{ghost}
{\cal K} \equiv \bar{\cal A} 
\bigg( 1 + \frac{\alpha \mu^2}{2f_2} \Big( \frac{{\cal A}_{1}}{\bar{\cal A}} 
+ \frac{a^2}{{\tilde k}^2+\alpha k_{\rm IR}^2} \frac{{\cal A}_{2}}{\bar{\cal A}} \Big)
\bigg) \,.
\end{equation}

Note that contributions from the kinetic couplings ${\cal A}_1$ and ${\cal A}_2$ to the sound speed \eqref{cs2} and the mass term \eqref{mass} appear after factoring out the total kinetic coefficient ${\cal K}$, expanding each term with small $\alpha\mu^2$, and then keeping only the leading-order terms. 

The conditions to have no ghost and no gradient instability are given by
\begin{eqnarray}\label{stability-conditions}
{\cal K} > 0 \,, \hspace{1cm} c_{\rm s}^2 ({\tilde k}) > 0 \,. 
\end{eqnarray}

Substituting $\zeta\propto e^{-i \omega t}$ in the total quadratic Lagrangian \eqref{Lagrangian-red}, we find the following dispersion relation for the mode $\zeta$
\begin{equation}\label{DR-ef}
\Big(\frac{\omega}{M}\Big)^2 = c_s^2(k) \Big(\frac{k}{aM}\Big)^2 + \alpha\mu^2 m^2 \,,
\end{equation}
where we have replaced the dimensionless momentum $\tilde k$ to the standard dimensionful momentum $k = \mu M {\tilde k}$. Note also that when we rewrite the sound speed squared in \eqref{DR-ef} in terms of the standard momentum $k$, the last term in Eq.~\eqref{cs2} would have a prefactor $\alpha$ without $\mu^2$. Therefore, this term is not suppressed and play the key role in our scenario to solve the issue of the strong coupling or gradient instability around the stealth solution. The dispersion relation~\eqref{DR-ef} reduces to the DHOST case Eq.~\eqref{DR-DHOST} for $\alpha=0$.

\section{Weakly-coupled stealth dark energy}\label{sec-stealth-DE}

In the previous sections we have studied the cosmological background and perturbations for the scordatura DHOST model \eqref{action-scordatura}. It is well known that this setup supports a stealth de Sitter solution which is strongly coupled in the absence of the scordatura term $\alpha=0$. In this section we show that the scordatura corrections make the setup weakly-coupled~\cite{Motohashi:2019ymr} and therefore we can construct a viable dark energy model in this framework.

The evolution equations~(\ref{Friedmann}) and (\ref{Raychuadhuri}), or equivalently Eqs.~\eqref{Friedmann-b} and \eqref{Raychuadhuri-b}, admit the following linearly time dependent solution 
\begin{equation}\label{stealth-sol}
\varphi({\tilde t}) = {\tilde t}\,, \hspace{1cm} {\mathrm x} = -1 \,,
\end{equation}
which after substituting in (\ref{Friedmann}) and (\ref{Raychuadhuri}) and some manipulations we find 
\begin{eqnarray}
f_0 +  3 h_{\rm dS}^2 \Big( 2 f_2 + \frac{3}{2} \alpha \mu^2 \Big) &=& 0 \,, \label{Stealth-Raychuadhuri}
\\
f_{0,{\mathrm x}} + 3 h_{\rm dS} ( 4 h_{\rm dS} f_{2,{\mathrm x}} - \mu f_{1,{\mathrm x}} ) &=& 0 \,. \label{Stealth-Friedmann}
\end{eqnarray}
Here, $h_{\rm dS}$ is the constant Hubble parameter that arises for the stealth solution (\ref{stealth-sol}). Note that Eq.~\eqref{Stealth-Friedmann} is equivalent to $\alpha_K =0$. From the above equations, we find 
\begin{equation}\label{Stealth-Hubble}
h_{\rm dS} = \sqrt{\frac{-f_0}{6 f_2+\frac{9}{2}\alpha\mu^2}} \,,
\end{equation}
where we note that all quantities are calculated for the stealth solution (\ref{stealth-sol}). Substituting the above result into (\ref{Stealth-Friedmann}), we find an algebraic equation for the constant ${\mathrm x}$. Note that for the stealth solution with $\ddot{\varphi}=0$, the Hubble parameters for the scale factors $a$ and $b$ coincide which can be seen from Eq. (\ref{Hubble-b}). Eqs.~(\ref{Stealth-Raychuadhuri}) and (\ref{Stealth-Friedmann}) coincide with those obtained in Ref.~\cite{Motohashi:2019ymr} and the stealth solution (\ref{Stealth-Hubble}) also coincides with the result that is obtained in Ref.~\cite{Crisostomi:2018bsp}.

Before studying the perturbations around the stealth solution, let us more carefully look at the stealth background solution \eqref{stealth-sol}. For the sake of simplicity, we focus on a subset of our model with $f_2=1/2$ and $f_1=0$. In this case, from \eqref{alpha-i}, we see $\alpha_H=0=\alpha_B$, and $\alpha_K= - \frac{{\rm x}}{3h^2} f_{0,{\rm x}}$ where we have used the fact that $h_b = h$ from \eqref{Hubble-b} for $\alpha_H=0$. Using these results in \eqref{j0-b}, we find $j^0=-2\dot{\varphi}f_{0,{\rm x}}$ and the equation of motion for the scalar field~\eqref{KG} implies $\big(a^3f_{0,{\rm x}}\dot{\varphi}\big)\dot{}=0$. This equation has a solution $f_{0,{\rm x}}\dot{\varphi}= c_1 a^{-3}$ with $c_1$ being an integration constant, which for expanding universe approaches to the attractor $f_{0,{\rm x}}\dot{\varphi}\to0$. There are two possibilities $\dot{\varphi}=0$ or $f_{0,{\rm x}}=0$. The first is nothing but the standard attractor solution for the massless scalar field while the latter is the stealth attractor which we presented in Eqs.~\eqref{stealth-sol} and \eqref{Stealth-Friedmann} with $\dot{\varphi}=\mbox{const.}$ and $\alpha_K= - \frac{{\rm x}}{3h^2} f_{0,{\rm x}}=0$ \cite{ArkaniHamed:2003uy}. For this attractor stealth solution with $\alpha_K=0=\alpha_H$ and $\ddot{\varphi}=0$, from Eq. \eqref{hb-dot}, we see that $\dot{h}_b = \dot{h} = 0$ which shows that the solution is exact de Sitter. Note that the above argument was based only on the equation of motion of the scalar field and it holds even after inclusion of matter as far as matter fields do not directly couple to the scalar field. For more general case of DHOST theories with $\alpha_H\neq0$ and $\alpha_B\neq0$, the stealth solution is still an attractor solution as it is shown in Ref. \cite{Crisostomi:2018bsp} by dynamical stability analysis.

Now, we look at the linear perturbations around the stealth solution \eqref{stealth-sol}. The gradient coefficient defined in (\ref{coefficients-AD-BD}) for the stealth solution and the sound speed square of the DHOST part that is defined in (\ref{DR-DHOST}) take the following simple forms
\begin{eqnarray}\label{cs2-D}
\bar{\cal B} = - 2 \frac{ \alpha_B - \alpha_H }{1+ \alpha_B} \,,\hspace{1cm}
\bar{c}_s^2 = \frac{1}{3} \frac{(1+\alpha_B)(\alpha_B-\alpha_H)}{\beta_K-(1+\alpha_B)^2} \,.
\end{eqnarray}

Substituting (\ref{alpha-i}) in the above results, we find explicit expressions in terms of the original functions
\begin{align}\label{Stealth-cs2}
\bar{c}_s^2 &= \frac{\mu f_{1,{\rm x}} \left(2 h_{\rm dS} \left(f_{2,{\rm x}} + f_2 \right)-\mu f_{1,{\rm x}} \right)}{4 f_2 f_{0,{\rm x}{\rm x}}+3 \mu ^2 f_{1,{\rm x}}^2+6 h_{\rm dS} 
	\mu  \left(f_{1,{\rm x}} \left(f_2-5 f_{2,{\rm x}} \right)-2 f_2 f_{1,{\rm x}{\rm x}}\right) 
	+ 48 h_{\rm dS} ^2 \left( f_2 f_{2,{\rm x}{\rm x}} + f_{2,{\rm x}}^2 \right)} \,, \\
\label{Stealth-kinetic}
\bar{\cal A} &= 2 \frac{ \left( 4 f_2 f_{0,{\rm x}{\rm x}}
+3 \mu^2 f_{1,{\rm x}}^2 + 6 \mu h_{\rm dS} \big( f_2 \big( f_{1,{\rm x}} - 2 f_{1,{\rm x}{\rm x}} \big) 
- 5 f_{1,{\rm x}} f_{2{\rm x}} \big) + 48 h_{\rm dS}^2 \left( f_2 f_{2,{\rm x}{\rm x}}
+ \left( f_{2,{\rm x}} \right)^2 \right) \right)}{ \big( \mu f_{1,{\rm x}}
-2 h_{\rm dS} \left( f_2 + f_{2,{\rm x}} \right)\big)^2} \,,
\end{align}
where we have also used (\ref{Stealth-Friedmann}) to eliminate $f_{0,{\mathrm x}}$. Demanding that $f_2>0$ to avoid the appearance of a ghost in the tensor sector, we find that the setup can be free from ghost and gradient instabilities for $0 < \mu f_{1,{\rm x}} < 2 h_{\rm dS}(f_2+f_{2,{\rm x}})$. Taking the fact $\mu\ll1$ into account, we see that the setup is always free from ghost while it provides small imaginary sound speed for $\mu f_{1,{\rm x}}<0$. This infrared instability is not malignant as it behaves as the Jeans instability \cite{ArkaniHamed:2003uy,ArkaniHamed:2005gu,Gumrukcuoglu:2016jbh}. If we consider the limit $\mu\to0$ (or $\alpha_B=\alpha_H$ that corresponds to the absence of kinetic braiding term), this instability disappears while the setup becomes strongly coupled as ${\bar c}_s\to0$. In the absence of the scordatura term ($\alpha=0$), this makes the strong coupling scale significantly lower than $M$ and thus diminishes the regime of validity of the EFT. With the scordatura term of order unity, on the other hand, the strong coupling scale is naturally raised to order unity in the unit of $M$.

We now take the effects of scordatura term into account. Around the stealth solution (\ref{Stealth-Raychuadhuri}) and (\ref{Stealth-Friedmann}), the Lagrangian coefficients defined in (\ref{Lagrangian-scordatura-red}) take the following forms
\begin{align}\label{coefficients-AS-BS-CS}
{\cal A}_{1} &= -\frac{9}{2 (1+\alpha_B)^4} \bigg[
\alpha _B^2 \left(15 \alpha _H^2+6 \alpha _H+7\right)
+2 \alpha _B \left(15 \alpha _H^2+6 \alpha _H \left(\beta _K+1\right)-2 \beta _K+7\right)
\nonumber \\
&~~~ +\,15 \alpha _H^2+6 \alpha _H \left(2 \beta_K+1\right)+2 \beta _K^2-4 \beta _K+7
\bigg] \,, 
\nonumber \\[5pt]
{\cal A}_{2} &= \frac{81}{2} h_{\rm dS}^2 
\bigg[ 1 - \frac{\beta_K}{(1+\alpha_B)^2} \bigg]^2 \,, \hspace{2cm} 
k_{\rm IR}^2 = - \frac{27}{8} \frac{\mu^2h_{\rm dS}^2a^4\beta_K}{f_2(1+\alpha_B)^2} \,,
\nonumber \\[10pt]
{\cal B}_{1} &= 
\frac{1}{2 \left(1+\alpha_B\right){}^3 \left((1+\alpha_B)^2 - \beta_K\right)}
\bigg[\alpha_B^2 \left(-18 \alpha_H^3+51 \alpha_H^2
+4 \alpha_H \left( 3 \beta_K+1\right)+8 \beta_K-41 \right)
\nonumber \\
&~~~ +\, \alpha_B^3 \left( 27 \alpha_H^2+6 \alpha_H-13 \right)
- \alpha_B \left(36 \alpha_H^3+3 \alpha_H^2 \left( 5 \beta_K-7 \right)
-2 \alpha_H \left(11 \beta_K-5\right)-29 \beta_K+43\right)
\nonumber \\
&~~~ -\,18 \alpha_H^3-3 \alpha_H^2 \left(5 \beta_K+1\right)
-2 \alpha_H \left(3 \beta_K^2-5 \beta_K+4\right)
-3 \left(2\beta_K^2-7 \beta_K+5\right)
\bigg]
\,, \nonumber \\[5pt]
{\cal B}_{2} &= \frac{1}{9 h_{\rm dS}^2} \bigg[ 
\frac{2\alpha_B^2 \left(2+3 \alpha_H\right)+\alpha_B 
\left(-3 \alpha_H^2+8 \alpha_H+7\right)-3
\alpha_H^2+\alpha_H \left(2-3 \beta _K\right)- 3 \beta_K+3}{\left(1+\alpha_B\right)
\left((1+\alpha_B)^2-\beta_K\right)}
\bigg]^2 \,, \nonumber \\[10pt]
{\cal M} &= - \frac{27}{2} h_{\rm dS}^2 \frac{ ( \alpha_B - 5) 
( 1 + \alpha_B )^2+3 ( 1 + \alpha_H ) \beta_K }{( 1 + \alpha_B )^3} \,.
\end{align}
Note that ${\cal A}_{2}>0$ and ${\cal B}_{2}>0$ independently of the explicit values of the functions $\alpha_i$ and $\beta_K$.

Substituting (\ref{Stealth-cs2}), (\ref{Stealth-kinetic}), (\ref{coefficients-AS-BS-CS}) into the dispersion relation \eqref{DR-ef} and then using (\ref{alpha-i}) to express the results in terms of the original functions $f_i$, we find
\begin{equation}\label{DR}
\Big(\frac{\omega}{M}\Big)^2 = \big( \bar{c}_s^2 
- \alpha \mu^2 \Gamma_1 \big) \Big(\frac{k}{aM}\Big)^2 
+ \alpha \Gamma_2 \Big(\frac{k}{aM}\Big)^4 + \alpha\mu^2m^2 
\end{equation}
with
\begin{eqnarray}\label{Gammas}
&&\Gamma_1 \equiv \frac{ 2 f_2 f_{0,{\rm x}{\rm x}} 
	+ 3 h_{\rm dS} ^2 \left(3 f_2^2
	+ 3 f_{2,{\rm x}}^2 + 2 f_2 \left(4 f_{2,{\rm x}{\rm x}}
	- f_{2,{\rm x}} \right) \right) }{8 f_2 \left( f_2 f_{0,{\rm x}{\rm x}} 
	+ 12 h_{\rm dS} ^2 \left( f_2 f_{1,{\rm x}{\rm x}} + f_{2,{\rm x}}^2\right)\right)}  \,,\nonumber \\ 
&&\Gamma_2 \equiv \frac{  \left( f_{2,{\rm x}} + f_2 \right)^2}{4 f_2 \left( f_2 f_{0,{\rm x}{\rm x}} 
	+ 12 h_{\rm dS} ^2 \left( f_2 f_{2,{\rm x}{\rm x}} 
	+ f_{2,{\rm x}}^2\right)\right)} \,, \nonumber \\ \label{m}
&&m^2 = 27 h_{\rm dS}^2 \, \frac{ \left( f_2 f_{0,{\rm x}{\rm x}} 
	+ 2 h_{\rm dS}^2 \left( f_2^2 + 4 f_{2,{\rm x}}^2 - f_2 \left( f_{2,{\rm x}}
	- 6 f_{2,{\rm x}{\rm x}} \right) \right) \right)}{ 8 f_2 \left( f_2 f_{0,{\rm x}{\rm x}}
	+12 h_{\rm dS}^2 \left( f_2 f_{2,{\rm x}{\rm x}} 
	+ f_{2,{\rm x}}^2\right)\right)} \,.
\end{eqnarray}

For $\mu\to0$, from \eqref{Stealth-cs2} and \eqref{m} we find ${\bar c}_s\to0$ and therefore the dispersion relation \eqref{DR-ef} simplifies to
\begin{equation}\label{DR-final}
	\frac{\omega}{M} \approx \sqrt{\alpha\Gamma_2} \Big(\frac{k}{aM}\Big)^2 \, .
\end{equation}
Note that the coefficient of the linear $k^2$ term in \eqref{DR} vanishes while the nonlinear $k^4$ term survives which gives nonzero contribution $\omega\propto k^2$.
This dependency agrees with that in the EFT of ghost condensation and is a key to make the system weakly-coupled~\cite{Motohashi:2019ymr}.

Finally, as a consistency check, we take the Minkowski limit $a\to1$ and $h_{\rm dS} \to 0$.
In this limit, the mass term vanishes and the dispersion relation (\ref{DR}) takes the following form
\begin{equation}\label{DR-Minkowski}
\Big(\frac{\omega}{M}\Big)^2 = - \left[ \mu^2 \frac{( f_{1,{\rm x}}^2 
+ \alpha f_{0,{\rm x}{\rm x}} )}{4 f_2 f_{0{\rm x}{\rm x}}} + \mathcal{O}(\mu^4) \right] \Big(\frac{k}{M}\Big)^2 
+ \alpha \frac{(f_2+f_{2,{\rm x}})^2}{4 f_2^2 f_{0,{\rm x}{\rm x}}} \Big(\frac{k}{M}\Big)^4 \,,
\end{equation}
which is in agreement with the result of \cite{Motohashi:2019ymr}.


\section{Coupling of the dark energy to dark matter}\label{sec-coupling-DM}

Having shown that our setup supports a weakly-coupled stealth dark energy, in this section, we investigate evolution of perturbations in the presence of the dark energy described by the scordatura DHOST scalar and the dark matter component. Since one can accommodate any background metric solution same as in GR in the presence of general matter component as a stealth background solution in DHOST theories~\cite{Takahashi:2020hso}, we will focus on the $\Lambda$CDM expansion as a background solution. To describe the matter sector, we consider a shift-symmetric k-essence field $\sigma$ that is minimally coupled to the gravity sector with the action
\begin{equation}\label{action-pf}
S_{\rm m} = \int d^4x \sqrt{-g} P(Y) \,, \hspace{1cm} Y = g^{\nu\eta} \sigma_\nu \sigma_\eta \,.
\end{equation}

Similar to the gravitational sector and in order to explicitly see the regime of validity of our scordatura EFT, we work with dimensionless coordinates (\ref{coordinets-ch}) and also define the following dimensionless quantities
\begin{equation}\label{dimensionless-k-essence}
\sigma = M_{P} \bar{\sigma} \,, \hspace{1cm} P\equiv{M^4}{\mathrm p}\,, 
\hspace{1cm} Y\equiv{M^4}{\mathrm y}  \,.
\end{equation}

The energy-momentum tensor of the shift-symmetric k-essence can be written as a perfect fluid energy-momentum tensor with the following energy density and pressure
\begin{equation}\label{rho-p}
\rho_{\rm m} \equiv 2 {\mathrm y} {\mathrm p}_{,{\mathrm y}} 
- {\mathrm p} \,, \hspace{1cm} p_{\rm m}  \equiv {\mathrm p} \,.
\end{equation}
Note that the energy density and pressure defined above are dimensionless.

Since the energy-momentum tensor for the shift-symmetric k-essence field can be modeled by a perfect fluid, we can translate all of our final results into the standard results of a perfect fluid working with appropriate variables.

We shall explore background dynamics in Section~\ref{ssec-debg} and the scalar perturbations in Section~\ref{ssec-dept}. In Section~\ref{ssec-degeff}, we shall find that the scordatura term is necessary to make the quasi-static limit well-defined. We shall then derive the evolution equation for the matter energy density contrast and obtain the gravitational coupling as well as corrections to the friction term in the presence of the scordatura term. We shall also clarify the scordatura contribution to the gravitational slip parameter.

\subsection{Background equations}
\label{ssec-debg}

The background equations can be easily found from the minimally coupled action $S_{\rm g} + S_{\rm m}$, where $S_{\rm g}$ and $S_{\rm m}$ are given by \eqref{action-scordatura} and \eqref{action-pf} respectively. For the stealth solution \eqref{stealth-sol} which in the absence of matter leads to the exact de Sitter configuration with the constant Hubble parameter \eqref{Stealth-Hubble}, the Friedmann equations in the presence of the perfect fluid take the forms
\begin{equation}\label{Friedmann-DE+DM-G}
6 f_2 h^2 = \rho_\Lambda + \rho_{\rm m}  \,, \hspace{1cm} 
- 2 f_2 \big( 2\dot{h} + 3 h^2 \big) =  p_\Lambda + p_{\rm m}  \,,
\end{equation}
where the energy density and pressure of the perfect fluid are given by the k-essence field which are defined in (\ref{rho-p}) while energy density and pressure of dark energy are already obtained as
\begin{equation}\label{rho-p-DE}
\rho_\Lambda = 6 f_2 h_{\rm dS}^2 = - p_\Lambda \,, \hspace{1cm} 
\frac{\Lambda}{M^2} \equiv 3 \mu^2 h_{\rm dS}^2 
= - \mu^2 f_0 \,,
\end{equation}
where, again, $f_0,f_2$ are evaluated at the stealth solution~\eqref{stealth-sol}, we have used Eq.~\eqref{Stealth-Hubble} and neglected the terms that are suppressed by the Planck scale.

In the case of pressureless dark matter $p_{\rm m}=0$, which we are interested in here, the conservation of energy momentum tensor gives
\begin{equation}\label{rhodot}
\dot{\rho}_{\rm m} + 3 h {\rho}_{\rm m} = 0 \,.
\end{equation}
This gives the solution ${\rho}_{\rm m} \sim a^{-3}$ and therefore the Friedmann equation becomes
\begin{equation}\label{Friedmann-DE+DM}
6 f_2 h^2 = \rho_\Lambda + \frac{\rho_{0{\rm m}}}{a^3} \,, 
\end{equation}
where $\rho_{0{\rm m}}={\rho}_{\rm m}({\tilde t}=0)$ is the energy density of the dark matter at the present time $a({\tilde t}=0)=1$. The solutions for the scale factor and the Hubble parameter are given by
\begin{equation}\label{hubble-DE+DM}
a({\tilde t}) = \Big(\frac{\rho_{0{\rm m}}}{ \rho_\Lambda}\Big)^{1/3} 
\sinh^{2/3}\bigg(\frac{\sqrt{ 3\rho_\Lambda}}{2}{\tilde t}\bigg) \,, \hspace{1cm}
h = h_{\rm dS} \coth\Big(\frac{3}{2}h_{\rm dS}{\tilde t}\Big) \,.
\end{equation}
At early time the above solution approaches to the matter dominated era with $a \sim {\tilde t}^{2/3}$ and $h \sim {\tilde t}^{-1}$ while it approaches the de Sitter solution at late time with $a \approx e^{h_{\rm dS}{\tilde t}}$ and $h \approx h_{\rm dS}$. 

Note that \eqref{hubble-DE+DM} is the stealth solution which is exactly the same as GR solution of the system with a cosmological constant plus pressureless dark matter at the background level, which is a special case of \cite{Takahashi:2020hso}. Therefore, at the background level, we cannot distinguish this model from the standard $\Lambda$CDM model. 
At the level of perturbations, however, these two setups are significantly different as we will see below.

\subsection{Perturbations}
\label{ssec-dept}

We have already performed the linear perturbation analysis for the gravitational sector which provides the dark energy component. Here, our aim is to redo the calculations in the presence of the minimally coupled dark matter. We thus take into account the scalar perturbations of the k-essence field
\begin{equation}\label{sigma-norm}
\sigma = M_{P} (\bar{\sigma} + \delta\sigma) \,,
\end{equation}
where ${\bar \sigma}$ and $\delta\sigma$ denote the background value and scalar perturbations respectively. Both of the ${\bar \sigma}$ and $\delta\sigma$ are dimensionless.

The observable quantity we are interested in is the matter energy density contrast.  Here we work with the following gauge-invariant quantity\footnote{Note that we cannot set $c_{\rm m} = 0$ from the beginning. We first perform the transformation (\ref{delta}) for finite $c_{\rm m}$ and impose $c_{\rm m} = 0$ at the end.}
\begin{equation}\label{delta}
\delta_{\rm m} = 3 \zeta - (1+3 \alpha_H c_{\rm m}^2) \frac{A}{c_{\rm m}^2} 
+ \frac{\dot{\delta\sigma}}{c_{\rm m}^2 \dot{\bar\sigma}} \,; \hspace{1cm}
c_{\rm m} ^2 \equiv \frac{{\mathrm p}_{,{\mathrm y}}}{{\mathrm p}_{,{\mathrm y}} 
+ 2 {\mathrm y} {\mathrm p}_{,{\mathrm y}{\mathrm y}}} \,,
\end{equation}
where $c_{\rm m}$ is the sound speed of the k-essence field.

The total second order action, which is the sum of the quadratic part of the gravity sector~(\ref{action-scordatura}) and the quadratic part of the matter sector (\ref{action-pf}), takes the form
\begin{equation}\label{action-SS}
S^{(2)}_{\rm g} + S^{(2)}_{\rm m} = \int d{\tilde t} d^3{\tilde x} M^4 
\Big[ {\tilde {\cal L}}^{(2)}_{\rm Dm}(\dot{\zeta},\zeta,A,B,\dot{\delta}_{\rm m},\delta_{\rm m}) 
+ \alpha \mu^2 {\tilde {\cal L}}_{\rm S}^{(2)}(\dot{\zeta},\zeta,\dot{A},A,B ) \Big] \,.
\end{equation}

The explicit form of the Lagrangian ${\tilde {\cal L}}^{(2)}_{\rm Dm}(\dot{\zeta},\zeta,A,B,\dot{\delta}_{\rm m},\delta_{\rm m})$, which represents the DHOST action minimally coupled to a perfect fluid, is obtained in the appendix~\ref{appendix-b} in Eq.~(\ref{Lagrangian-DHOST-m-a}). For the dark matter with $p_{\rm m}= 0$ (or equivalently $w_{\rm m}=0$ where $w_{\rm m}=p_{\rm m}/\rho_{\rm m}$ is the equation of state parameter) and $c_{\rm m}= 0$, it simplifies to
\begin{eqnarray}\label{Lagrangian-DHOST-m}
{\tilde {\cal L}}_{\rm Dm}^{(2)} =
{\tilde {\cal L}}_{\rm D}^{(2)} 
+ \frac{1}{2} a^3 \rho_{\rm m} \Big( 
\frac{a^2}{{\tilde k}^2} \dot{\delta}_{\rm m}^2 
+ 2 B \dot{\delta}_{\rm m} - 2 A \delta_{\rm m}
+ \frac{{\tilde k}^2}{a^2} B^2 + 6 A \zeta 
- 3 \beta_{\rm m} A^2 \Big)\,,
\end{eqnarray}
where ${\tilde {\cal L}}_{\rm D}^{(2)}$ is defined in Eq.~\eqref{Lagrangian-DHOST} and we have also defined another second order dimensionless parameter 
\begin{eqnarray}\label{beta-v}
\beta_{\rm m} \equiv \frac{ 2 \beta_H - 5 \alpha _H^2 - \alpha _H }{1-3 \alpha _H} \,.
\end{eqnarray}

The explicit form of the quadratic Lagrangian for the scordatura part ${\tilde {\cal L}}_{\rm S}^{(2)}(\dot{\zeta},\zeta,\dot{A},A,B )$ is still given by (\ref{Lagrangian-scordatura}) since the matter is not directly coupled to the scordatura term and also we did not use the background equations in obtaining (\ref{Lagrangian-scordatura}).

\subsection{Effective gravitational coupling in quasi-static limit}
\label{ssec-degeff}

Having obtained the quadratic action for the linear perturbations of the system of stealth dark energy in the presence of dark matter minimally coupled to gravity, now we look at the formation of structures by adopting the so-called quasi-static approximation. This approximation is justified for the modes deep inside the horizon $k \gg (aH)/c_s$ where $c_s$ is the dark energy sound speed \cite{Bellini:2014fua,Sawicki:2015zya}. In the absence of the dark matter, we have found the sound speed of the dark energy which is given by Eq.~\eqref{cs2}. In the presence of dark matter, the sound speed changes through the integrating out of the non-dynamical variables. However, these corrections are suppressed by the Planck scale and, therefore, Eq.~\eqref{cs2} still determines the sound speed of the dark energy. Far from the stealth solution, the dominant contribution is given by the DHOST part and we have $c_s \approx {\bar c}_s$ of order unity and the condition $k \gg (aH)/{\bar c}_s$ can then be satisfied \cite{Gleyzes:2015pma}. For the stealth solution, as it is clear from \eqref{Stealth-cs2}, ${\bar c}_s\to0$ and as we have shown in Eq.~\eqref{DR-final}, the dominant contribution to the sound speed is given by the scordatura term as $c_s(k)\approx \sqrt{\alpha\Gamma_2} (k/aM)$. Therefore, for the stealth solution, existence of the scordatura term is essential to justify the quasi-static approximation. Otherwise, the condition $k \gg (aH)/c_s$ will be violated for $\alpha\to0$.

Varying the quadratic Lagrangian ${\tilde {\cal L}}^{(2)}_{\rm Dm} + \alpha \mu^2 {\tilde {\cal L}}_{\rm S}^{(2)}$ with respect to $B,A$, and $\zeta$, where ${\tilde {\cal L}}^{(2)}_{\rm Dm}$ and ${\tilde {\cal L}}_{\rm S}^{(2)}$ are given by \eqref{Lagrangian-DHOST-m} and \eqref{Lagrangian-scordatura}, we find the following equations of motion respectively
\begin{align}\label{EOM-B}
& h (1+\alpha_B) A- \dot{\zeta} 
+ \epsilon_{\rm m} h^2 \Big( \frac{a^2}{{\tilde k}^2}\dot{\delta}_{\rm m} + B \Big) = 
\frac{\alpha\mu^2}{4 f_2 a} \bigg[
- {\bar n}_{13} \dot{\zeta} - {\bar n}_{23} \dot{A} + {\bar m}_{23} A 
- ({\bar m}_{33}+ {\bar m}_{33{\rm s}} {\tilde k}^2) B \bigg]
 \,, \\ 
& 3 h \left( (1+\alpha_B) \dot{\zeta}-h \beta_K A \right)
+  \frac{{\tilde k}^2}{a^2} \big( h (1+\alpha_B) B + (1+\alpha_H) \zeta \big) -
\epsilon_{\rm m} h^2 \big( \delta_{\rm m} 
- 3\zeta  + 3  \beta_{\rm m} A \big) 
\nonumber \\ \label{EOM-A}
& \hspace{2cm} = \frac{\alpha\mu^2 }{4 f_2 a^3} \bigg[
\frac{d}{dt} \Big( {\bar k}_{22} \dot{A} + {\bar k}_{12} \dot{\zeta} 
+ {\bar n}_{23} {\tilde k}^2 B \Big) - {\bar n}_{12} \dot{\zeta} 
+ {\bar m}_{12} \zeta + {\bar m}_{22} A + {\bar m}_{23} {\tilde k}^2 B 
\bigg] , \\
& \frac{1}{a^3} \frac{d}{dt} \bigg[ a^3 \Big(3\dot{\zeta} - (1+\alpha_B) h A 
+ \frac{{\tilde k}^2}{a^2} B \Big) \bigg]
+ \frac{{\tilde k}^2}{a^2} \zeta 
+ (1+\alpha_H)  \frac{{\tilde k}^2}{a^2} A + 3 \epsilon_{\rm m} h^2 A 
\nonumber \\ \label{EOM-zeta}
&\hspace{2cm} = - \frac{\alpha\mu^2}{4 f_2 a^3} 
 \bigg[
\frac{d}{dt} \Big( {\bar k}_{11} \dot{\zeta} + {\bar k}_{12} \dot{A} 
+ {\bar n}_{12} A + {\bar n}_{13} {\tilde k}^2 B \Big) 
+ {\bar m}_{11} \zeta + {\bar m}_{12} A \bigg] \,.
\end{align}
In obtaining the above equations of motion, we have used the fact that $h_b=h$ for the stealth solution and also defined the dimensionless quantity
\begin{equation}\label{epsilon-m}
\epsilon_{\rm m} \equiv - \frac{\dot{h}}{h^2} = \frac{\rho_{\rm m}}{4 f_2 h^2} \,,
\end{equation}
where in the last step we have used the background equation $\dot{h} = - \frac{\rho_{\rm m}}{4 f_2}$.

Variation with respect to $\delta_{\rm m}$ gives the equation of motion for the matter energy density contrast
\begin{equation}\label{EOM-deltam}
\ddot{\delta}_{\rm m} + 2 h \dot{\delta}_{\rm m} = - \frac{{\tilde k}^2}{a^2} \big( A+ \dot{B} \big) \,.
\end{equation}

Solving Eqs.~\eqref{EOM-B} and \eqref{EOM-A} in the limit $\mu\to0$ for the non-dynamical modes $A$ and $B$ and then substituting the result into \eqref{EOM-zeta} and \eqref{EOM-deltam}, after some manipulations, we find second order equations of motion $\ddot{\zeta} = \ddot{\zeta}(\dot{\zeta},\zeta,\dot{\delta}_{\rm m},\delta_{\rm m})$ and $\ddot{\delta}_{\rm m} = \ddot{\delta}_{\rm m}(\dot{\zeta},\zeta,\dot{\delta}_{\rm m},\delta_{\rm m})$ which have complicated forms. In this respect, the dynamics of the two dynamical modes $\zeta$ and $\delta_{\rm m}$ is completely determined by the equations \eqref{EOM-B}, \eqref{EOM-A}, \eqref{EOM-zeta} and \eqref{EOM-deltam}. We can then find master equation for $\delta_{\rm m}$ which is fourth order in time derivative and then taking the quasi-static limit to neglect the higher-order time derivatives. Note that if we do not work in the limit $\mu\to0$, we also have another second order equation of motion for the ghost mode $A$ and the master equation for $\delta_{\rm m}$ will be sixth order, which makes the calculations very complicated. Alternatively, we can adopt the quasi-static approximation before integrating out the non-dynamical fields as in \cite{Tsujikawa:2007gd,DeFelice:2010as,DeFelice:2011hq,DeFelice:2011aa,Gleyzes:2015pma,DAmico:2016ntq,Kase:2020hst}, which is much easier as we will show below. In this limit, Eqs.~\eqref{EOM-B}, \eqref{EOM-A}, and \eqref{EOM-zeta} simplify to
\begin{align}\label{EOM-B-ST-QS}
& \dot{\zeta} - h (1+\alpha_B) A - \epsilon_{\rm m} h^2 B 
+ \frac{\alpha \mu^2 }{4 f_2} \frac{{\tilde k}^2}{a^2} B
= \epsilon_{\rm m} h^2 \frac{a^2}{{\tilde k}^2} \dot{\delta}_{\rm m} \,, \\ \label{EOM-A-ST-QS}
& (1+\alpha_H) \zeta + h (1+\alpha_B) B =
\epsilon_{\rm m} h^2 \frac{a^2}{{\tilde k}^2} \delta_{\rm m} \,, \\ \label{EOM-zeta-ST-QS}
& \zeta + h B + \dot{B} + (1+\alpha_H) A = 0 \,,
\end{align}
where, similarly to the background equations, we have neglected the scale-independent terms that are proportional to $\alpha$ since they are all suppressed by the Planck scale. 

Now, our task is to find $A$ and $B$ in terms of $\delta_{\rm m}$ and $\dot{\delta}_{\rm m}$ from Eqs.~\eqref{EOM-B-ST-QS}, \eqref{EOM-A-ST-QS}, and \eqref{EOM-zeta-ST-QS}, and then substituting the results into Eq.~\eqref{EOM-deltam} to find the equation of motion for the linear energy density contrast. The large-scale observable quantities are defined in terms of the Bardeen potentials
\begin{equation}\label{Bardeen-potentials}
\Phi = \psi + hB = \zeta+hB - \alpha_H A \,, \hspace{1.5cm} \Psi = \dot{B} + A \,,
\end{equation}
and we therefore rewrite our results in terms of these gauge-invariant variables.

Rewriting Eq.~\eqref{EOM-zeta-ST-QS} in terms of Bardeen potentials \eqref{Bardeen-potentials} and solving it for $A$ gives
\begin{equation}\label{Eq8}
A = -\frac{\Phi +\Psi }{2 \alpha_H} \,.
\end{equation}
Substituting \eqref{Bardeen-potentials} and \eqref{Eq8} in \eqref{EOM-A-ST-QS} yields
\begin{equation}\label{Eq9}
 ( 1 + \alpha_H ) (\Phi -\Psi ) + 2 (\alpha _B-\alpha_H ) h B
= 2 \epsilon_{\rm m} h^2 \frac{a^2}{{\tilde k}^2} \delta_{\rm m}  \,.
\end{equation}
Taking a time derivative of \eqref{EOM-A-ST-QS}, erasing $\dot{\zeta}$ by using \eqref{EOM-B-ST-QS}, and rewriting the results in terms of the Bardeen potentials, we find
\begin{equation}\label{Eq20}
(1+\alpha_B) (\Phi -\Psi ) + (1+\alpha_H) \frac{\alpha\mu^2}{2f_2 h}\frac{ {\tilde k}^2}{a^2} B
= 2 \epsilon_{\rm m} h \frac{a^2}{{\tilde k}^2} \big( h \delta_{\rm m}+\alpha _H \dot{\delta}_{\rm m} \big) \,,
\end{equation}
where we have used the relation $\dot{\alpha}_{B} = - \frac{\dot{h}}{h}(\alpha_B-\alpha_H) = \epsilon_{\rm m} h (\alpha_B-\alpha_H)$ which is valid around the stealth solution \eqref{hubble-DE+DM}.

From Eqs.~\eqref{Eq9} and \eqref{Eq20} we find
\begin{equation}\label{solB}
B = \frac{\epsilon_{\rm m}}{{\rm d}} \frac{a^2}{{\tilde k}^2} \,
\Big[ (\alpha_B-\alpha_H) h \delta_{\rm m} 
- \alpha_H (1+\alpha_H) \dot{\delta}_{\rm m} \Big] \,,
\end{equation}
and
\begin{equation}\label{Weyl-potential}
\frac{\Psi-\Phi}{2} = - \frac{\epsilon_{\rm m}h}{{\rm d}} \frac{a^2}{{\tilde k}^2} 
\bigg[ (\alpha_B - \alpha_H) 
\big( h\delta_{\rm m} + \alpha_H \dot{\delta}_{\rm m} \big) 
- (1+\alpha_H) \frac{\alpha \mu^2}{4 f_2 h} \frac{{\tilde k}^2}{a^2} \delta_{\rm m} \bigg] \,,
\end{equation}
where we have defined
\begin{equation}\label{d-def}
{\rm d} \equiv \left(1+\alpha_B\right) (\alpha_B-\alpha_H) 
- \frac{\alpha \mu ^2}{4 f_2 h^2} \frac{ {\tilde k}^2}{a^2}(1+\alpha_H){}^2 \,.
\end{equation}
The quantity $(\Psi - \Phi)/2$ in the left hand side of \eqref{Weyl-potential} is called the Weyl potential which measures the anisotropic stress.

Substituting \eqref{solB} in \eqref{EOM-A-ST-QS}, we find the dynamical field $\zeta$ in terms of the matter energy density contrast and its first time derivative as 
\begin{equation}\label{solxi}
\zeta = \frac{\epsilon_{\rm m}h}{{\rm d}} \frac{a^2}{{\tilde k}^2} 
\Big( \alpha_H (1+\alpha_B) \dot{\delta}_{\rm m} 
- (1+\alpha_H) \frac{\alpha \mu^2}{4 f_2 h} \frac{ {\tilde k}^2}{a^2} \delta_{\rm m} \Big) \,.
\end{equation}

Taking time derivative of the above relation and then using \eqref{EOM-deltam} to eliminate $\ddot{\delta}_{\rm m}$, we find $\dot{\zeta} = \dot{\zeta}(\Psi,\delta_{\rm m},\dot{\delta}_{\rm m})$. Substituting this result together with \eqref{Eq8} and \eqref{solB} in \eqref{EOM-B-ST-QS}, we find
\begin{align}\label{Eq21}
& {\rm d} (\Psi + \Phi)
- 2 \alpha_H^2 \epsilon_{\rm m} \Psi = 
2 \alpha_H \frac{ \epsilon_{\rm m} h}{ {\rm d} } \frac{a^2}{{\tilde k}^2}
\bigg[ \Big( {\rm d} \alpha
_B+\alpha _H \left(\alpha _B-\alpha _H\right) \big(2 (1+\alpha_B)-(1+\alpha_H) \epsilon_{\rm m}\big)\Big) \dot{\delta}_{\rm m}
 \\ \nonumber
& 
+ \Big( {\rm d} (1-\alpha_B + 2 \alpha_H) (\alpha _B-\alpha_H)
-(1+\alpha_H) \left(\alpha _B-\alpha _H\right){}^2
\big(2(1+ \alpha_B)-(1+\alpha_H) \epsilon_{\rm m}\big)
+{\rm d}^2 \Big) \frac{h \delta_{\rm m}}{(1+\alpha_H){}^2}
\bigg] .
\end{align}

Now, from Eqs.~\eqref{Weyl-potential} and \eqref{Eq21}, we find the Bardeen potentials completely in terms of the matter energy density contrast 
\begin{align}\label{sol-Phi-Psi}
&\Psi = \frac{h \epsilon_{\rm m}}{ {\rm d} ({\rm d} - \alpha_H^2 \epsilon_{\rm m} )} \frac{{\tilde k}^2}{a^2}
\Bigg[
\Big( (\alpha_B-\alpha_H) \big(2 ( 1+ \alpha_B) -(1+\alpha_H) \epsilon_{\rm m} \big)+ {\rm d} \Big) 
\alpha_H^2 \dot{\delta}_{\rm m}
 \\ \nonumber
&
+ \Big( {\rm d} \left(\alpha _B-\alpha _H\right) (\alpha_B+\alpha_H^2)
- \alpha_H (1+\alpha_H) (\alpha_B-\alpha_H){}^2 \left(2 (1+\alpha_B) - (1+\alpha_H) \epsilon_{\rm m}\right) + {\rm d}^2 \Big) \frac{h \delta_{\rm m}}{(1+\alpha_H){}^2}
\Bigg] , \\ \label{sol-Phi}
& \Phi =  \frac{h \epsilon_{\rm m}}{ {\rm d} ({\rm d} - \alpha_H^2 \epsilon_{\rm m} )} \frac{{\tilde k}^2}{a^2}
\Bigg[
\Big( {\rm d}
\left(2 \alpha _B-\alpha _H\right)+\alpha _H \left(\alpha _B-\alpha _H\right)
\big(2 (1+\alpha_B) - (1+3 \alpha_H) \epsilon_{\rm m}\big) \Big) \alpha_H \dot{\delta}_{\rm m}
 \\ \nonumber
& \hspace{3.5cm} - \bigg({\rm d} \left(\alpha_B^2 (1+2 \alpha_H)-\alpha _B \alpha_H (3+5 \alpha_H)
+\alpha_H^2 \big(2+ 3 \alpha_H+2 (1+\alpha_H) \epsilon_{\rm m} \big)\right) 
\\ \nonumber
& 
\hspace{2.5cm} + \alpha_H (1+\alpha_H) \left(\alpha_B-\alpha_H\right){}^2 
\big(2 (1+\alpha_B) - (1+3\alpha_H) \epsilon_{\rm m} \big)- {\rm d}^2 
(1+2 \alpha_H)\bigg) \frac{h \delta_{\rm m}}{(1+\alpha_H){}^2}
\Bigg] .
\end{align}

Rewriting \eqref{EOM-deltam} in terms of the Bardeen potential $\Psi$ defined in \eqref{Bardeen-potentials}, and then substituting the above solution, we find
\begin{equation}\label{EOM-deltam-f}
\ddot{\delta}_{\rm m} + ( 2 + {\gamma} ) h \dot{\delta}_{\rm m} 
= 4 \pi {G}_{\rm eff} \rho_{\rm m} {\delta}_{\rm m} \,,
\end{equation}
where we have defined the effective gravitational coupling for the matter energy density contrast
\begin{eqnarray}\label{G_eff}
{G}_{\rm eff} \equiv \frac{1}{16 \pi f_2}\, \Bigg[
\frac{{\rm d}^2 - {\rm d} \left(\alpha _B-\alpha _H\right) (\alpha_B+\alpha_H^2)
+\alpha_H (1+\alpha_H) (\alpha_B-\alpha_H){}^2 
\big(2 (1+\alpha_B) - (1+\alpha_H) \epsilon_{\rm m}\big)}{
{\rm d} (1+\alpha_H){}^2 \left({\rm d}-\alpha_H^2 \epsilon_{\rm m}\right)}
\Bigg] , 
\end{eqnarray} 
and also the correction to the friction term
\begin{eqnarray}
\gamma \equiv
\frac{\alpha_H^2 \epsilon_{\rm m}}{{\rm d} ({\rm d}-\alpha_H^2 \epsilon_{\rm m} )} 
\Big( {\rm d} + (\alpha_B-\alpha_H) 
\big(2(1+\alpha_B)-(1+\alpha_H) \epsilon_{\rm m} \big) \Big)
 \,. \label{gamma}
\end{eqnarray}
As we can see, the effects of the perturbations in the gravity sector are first to modify the effective gravitational coupling \cite{Tsujikawa:2007gd,DeFelice:2010as,DeFelice:2011hq,DeFelice:2011aa,Gleyzes:2015pma,DAmico:2016ntq,Kase:2020hst} and second to change the friction term \cite{Kobayashi:2014ida}. Both of these effects affect the formation of the large-scale structures in the universe.

In the absence of the scordatura corrections, Eqs.~\eqref{G_eff} and \eqref{gamma} reduce to
\begin{eqnarray}\label{G_eff-DHOST}
{G}_{\rm eff}\big{|}_{\alpha=0} &=& \frac{1}{16 \pi f_2} \frac{\alpha_B-\alpha_H}{1+\alpha_B} \, 
	\frac{ 1 + \alpha_B - \alpha_H \epsilon_{\rm m} }{
		(1+\alpha_B) \left(\alpha _B-\alpha _H\right)-\alpha _H^2 \epsilon_{\rm m}} \,,
	\\
\label{gamma-DHOST}
\gamma|_{\alpha=0} &=&
	\frac{\alpha_H^2 \epsilon_{\rm m}}{1+\alpha_B}
	\frac{3(1+\alpha_B) - (1+\alpha_H) \epsilon_{\rm m}}{(1+\alpha_B)(\alpha_B-\alpha_H)-\alpha_H^2\epsilon_{\rm m}}
	\,,
\end{eqnarray}
which tell us an important role of the scordatura term for the well-definedness of the quasi-static limit.
We note that in the limit $\alpha_B \to \alpha_H$, which corresponds to the absence of the kinetic braiding term in the action~(\ref{action-scordatura}) as it is clear from \eqref{alpha-i}, the effective gravitational coupling~\eqref{G_eff-DHOST} vanishes. However, we are not allowed to take this limit in the absence of the scordatura term. This can be understood if we note that in the absence of the scordatura term $\alpha=0$, the dark energy sound speed reduces to ${\bar c}_s$ given by \eqref{cs2-D} which vanishes for $\alpha_B \to \alpha_H$. Not only it means infinite strong coupling, but also it is clear that the quasi-static limit $k \gg (aH)/{\bar c}_s$ is ill-defined in the limit ${\bar c}_s\to0$. However, in the presence of the scordatura term, the condition $k \gg (aH)/{c}_s$ holds even for ${\bar c}_s\to0$ and therefore we can safely take the limit $\alpha_B \to \alpha_H$ in \eqref{G_eff}. On the other hand, for the friction correction~\eqref{gamma-DHOST} for the DHOST subset, we can take the limit $\alpha_B \to \alpha_H$, which gives $\gamma = -3 + \epsilon_{\rm m}$. However, the scale-dependent corrections due to the scordatura term in \eqref{gamma} are not suppressed and we should take into account their effects.

Another observable quantity is the gravitational slip parameter that is defined as
\begin{eqnarray}\label{slip-p}
	\eta = - \frac{\Phi}{\Psi}
	\,, 
\end{eqnarray}
which after substituting \eqref{sol-Phi-Psi} and \eqref{sol-Phi} can be determined up to the ratio $\dot{\delta}_{\rm m}/(h\delta_{\rm m})$.

In summary, in this section we have shown that the scordatura prescription first allows us to define the quasi-static limit and second gives scale-dependent corrections to the physical quantities such as the Weyl potential \eqref{Weyl-potential}, the effective gravitational coupling \eqref{G_eff}, the correction~\eqref{gamma} to the friction term of the dark matter energy density contrast, and the slip parameter \eqref{slip-p}. The scale-dependent corrections to the effective gravitational coupling \eqref{G_eff} from the scordatura term are necessary to make this quantity well-defined in the quasi-static limit. Moreover, these scale-dependent corrections involve the scordatura effects and hence make our stealth dark energy model to be observationally different than not only the standard $\Lambda$CDM model but also many other dark energy models that are based on ghost-free modified gravity theories without the scordatura. It is interesting to perform more detailed analysis of the observational signatures of the scordatura, but it is beyond the scope of this work.

\section{Summary and conclusions}\label{sec-summary}

The stealth solution is an interesting class of exact solutions in scalar-tensor theories as its whole effects at the background level are to shift the cosmological constant and hence the background metric takes the same form as in GR.  The differences between the original background solutions in GR and the corresponding stealth background solutions show up at the level of perturbations, for which either strong coupling or gradient instability is inevitable for asymptotically flat or de Sitter stealth solution in any scalar-tensor theories possessing second-order equations of motion for perturbations~\cite{Motohashi:2019ymr}.  A universal prescription to resolve this problem is to introduce a controlled detuning of the degeneracy condition dubbed scordatura~\cite{Motohashi:2019ymr}, which renders the perturbations weakly coupled with the cost of a benign apparent Ostrogradsky ghost above the EFT cutoff scale.

In this paper we constructed a stealth dark energy model based on the weakly-coupled stealth de Sitter solution in the scordatura scenario.  We adopted the background metric same as in the standard $\Lambda$CDM model, and the linearly time-dependent scalar field profile with a constant kinetic term.  We investigated the effects of the scalar field perturbations of the dark energy on the dark matter perturbations which are responsible for the formation of the large-scale structure.  By virtue of the scordatura theory, we obtained non-vanishing, scale-dependent effective sound speed $c_{s}(k)$ for the scalar perturbations, which is a key to avoid the gradient instability and strong coupling.  Further, we clarified that the scordatura mechanism is also necessary to make the quasi-static limit $k \gg (aH)/c_s$ well-defined, which has been commonly adopted to study the evolution of the linear perturbation modes deep inside the sound horizon.  We thus conclude that the subhorizon observables are inevitably affected by the scordatura. 

We obtained the linear equation of motion for the subhorizon evolution of the dark matter energy density contrast, from which we found the corresponding effective gravitational coupling as well as a correction to the friction term.  We also found scale-dependent corrections due to the scordatura term to the Weyl potential and the gravitational slip parameter.  Although the scordatura is necessary to make the quasi-static limit well-defined, whether the contribution from the scordatura is sub/dominant in observables should be studied on case-by-case basis in comparison with the contribution from the ghost-free part of a scalar-tensor theory.  We highlighted that the most drastic case is a class of theories with $\alpha_B=\alpha_H$, i.e.\ a class without the kinetic braiding term. In such a class, in the absence of the scordatura the effective gravitational coupling appears to vanish.  Such a result is untrustable since in the absence of the scordatura the sound speed vanishes and hence the scalar field perturbation is infinitely strongly coupled and the quasi-static limit is ill-defined.  Hence, for this class, it is crucial to take the scordatura into account, and one can directly hear the sound of the scordatura by focusing on the evolution of perturbations.

The Ostrogradsky ghost-free higher-derivative theories serve as a unified framework to describe physics at low energy regime.  However, unless the degeneracy condition is protected by a fundamental symmetry, it will be eventually broken by quantum corrections.  If such a violation is of $\mathcal{O}(1)$ or less in the unit of the EFT cutoff scale, it resolves theoretical issues and provides rich phenomenology while retaining the apparent Ostrogradsky ghosts above the cutoff.  Namely, the scordatura resolves the strong coupling and the gradient instability for the scalar field perturbation around stealth solutions, recovers the generalized second law of black hole thermodynamics, and is necessary to make the quasi-static regime well-defined.  Using the results of the present paper, it is important to clarify the effects of the scordatura on observables and evaluate its significance compared to the one from the pure DHOST part of scalar-tensor theories.  Previous estimations of the subhorizon evolution of matter density contrast in modified gravity in the literature need to be revisited by taking into account the scordatura effect.  We leave it as a future work.

\vspace{.5cm}
{\bf Acknowledgments:} 
We would like to thank Andrei Lazanu and Philippe Brax for pointing out important typos. M.A.G.\ acknowledges the xTras package~\cite{Nutma:2013zea} which was used for tensorial calculations. The work of M.A.G. was supported by Japan Society for the Promotion of Science (JSPS) Grants-in-Aid for international research fellow No. 19F19313. H.M.\ was supported by JSPS Grants-in-Aid for Scientific Research (KAKENHI) No.\ 18K13565. The work of S.M.\ was supported by JSPS KAKENHI No.\ 17H02890, No.\ 17H06359, and by World Premier International Research Center Initiative, MEXT, Japan. 

\vspace{0.2cm}

\appendix

\section{Dimensionless scalar perturbations}\label{appendix}
\setcounter{equation}{0}
\renewcommand{\theequation}{A\arabic{equation}}

In order to explicitly see which higher-derivative term is suppressed from the EFT point of view, it is better to work with dimensionless quantities. In this appendix, considering the symmetries of the background, we systematically find the appropriate coordinates to redefine all background and perturbation quantities to their dimensionless quantities.

We consider linear scalar perturbations around the spatially flat FLRW metric. The metric in the comoving gauge in the (dimensionful) coordinates~$(t,x^i)$ takes the form
\begin{equation}\label{metric-FRW}
ds^2 = - ( 1 + 2 A ) dt^2 
+ 2 \partial_i {\bar B} dt dx^i 
+ a^2 ( 1 + 2 \psi ) \delta_{ij} dx^i dx^j \,,
\end{equation}
where $a(t)$ is the scale factor, $A(t,x^i), {\bar B}(t,x^i), \psi(t,x^i)$ are scalar perturbations, $t$ denotes the cosmic time, and $\partial_i$ denotes a spatial derivative with respect to the coordinates $x^i$ with $i=1,2,3$.

Let us first look at the background where the scalar perturbations in (\ref{metric-FRW}) are absent. We need to look at the Ricci scalar of the background which includes the following term
\begin{equation}\label{RicciS-BG}
R^{(0)} \supset \Big(\frac{\partial_t a}{a}\Big)^2 \,.
\end{equation}
The Ricci scalar has dimension $[R] = M^4/M_P^2$ and thus we find the following dimensionless time coordinate
\begin{equation}\label{time-def}
{\tilde t} \equiv \mu M t \,; \hspace{1cm} 
\mu\equiv {M/M_P}\,.
\end{equation}
From the background analysis we cannot find the appropriate spatial coordinates which is the consequence of the homogeneity of the background metric.

For the linear perturbations, the Ricci scalar includes the following terms
\begin{equation}\label{RicciS-Perturbations}
R^{(1)} \supset \Big\{ 
\Big(\frac{\partial_{t} a}{a}\Big)^2 {A} \,, \frac{\partial_i^2{A}}{a^2} \,,
\partial_{ t}^2{ \psi} \,, \frac{\partial_i^2{\psi}}{a^2} \,,
\partial_i^2\partial_{ t}{\bar B} 
\Big\} \,.
\end{equation}
From the terms that include $\partial_i^2{A}$ and $\partial_i^2{\psi}$ we realize the dimensionless spatial coordinates ${\tilde x}^i$ as
\begin{equation}\label{spatial-def}
{\tilde x}^i \equiv \mu M { x}^i \,.
\end{equation}
Looking at the last term in (\ref{RicciS-Perturbations}), we find the following dimensionless scalar mode
\begin{equation}\label{B-def}
B \equiv \mu M {\bar B} \,.
\end{equation}

Now, substituting (\ref{time-def}), (\ref{spatial-def}) and (\ref{B-def}) in (\ref{metric-FRW}), we find the metric in terms of the dimensionless coordinates~$(\tilde t,\tilde x^i)$ and scalar modes as follows
\begin{equation}\label{metric-FRW-Dimensionless}
ds^2 = \frac{M_P^2}{M^4} \Big[ - ( 1 + 2 A ) d{\tilde t}^2 
+ 2 {\tilde \partial}_i B d{\tilde t} d{\tilde x}^i 
+ a^2 ( 1 + 2 \psi ) \delta_{ij} d{\tilde x}^i d{\tilde x}^j \Big] \,,
\end{equation}
which we adopt throughout the present paper to study background and perturbations analysis.

In the same manner, we need to define appropriate dimensionless quantities in the dark energy and dark matter sectors which only include scalar fields $\phi$ and $\sigma$ in our model. Let us explain the logic for $\phi$ as the same holds for $\sigma$. The kinetic term of the scalar field has the dimension of $[X]=M^4$ and we define its dimensionless counterpart as ${\mathrm x}\equiv {X/M^4}$. The background value for ${\mathrm x}$ in the dimensionless coordinates defined in (\ref{metric-FRW-Dimensionless}) is ${\mathrm x}({\tilde t}) = - M_P^{-2} (\partial_{\tilde t}\phi)^2$, and, therefore, we define the dimensionless quantity for the scalar field $\varphi\equiv M_P^{-1}\phi$.

\section{Quadratic action in the presence of a minimally coupled perfect fluid}\label{appendix-b}
\setcounter{equation}{0}
\renewcommand{\theequation}{B\arabic{equation}}

In this appendix, we obtain the total quadratic action when our model is minimally coupled to a perfect fluid. For the sake of simplicity, we work with a shift symmetric k-essence field defined by the action (\ref{action-pf}). We translate the result in terms of the gauge-invariant energy density and, therefore, all the results can be considered for a general perfect fluid.

The total second-order action is the sum of the quadratic part of the gravity sector (\ref{action-scordatura}) and the quadratic part of the matter sector (\ref{action-pf}) which takes the form 
\begin{equation}\label{action-SS-a}
S^{(2)}_{\rm g} + S^{(2)}_{\rm m} = \int d{\tilde t} d^3{\tilde x} M^4 
\Big[ {\tilde {\cal L}}^{(2)}_{\rm Dm}(\dot{\zeta},\zeta,A,B,\dot{\delta\sigma},\delta\sigma) 
+ \alpha \mu^2 {\tilde {\cal L}}_{\rm S}^{(2)}(\dot{\zeta},\zeta,\dot{A},A,B ) \Big] \,,
\end{equation}
where $\delta\sigma$ is the perturbation of the k-essence field defined in (\ref{sigma-norm}). In the above quadratic action, ${\tilde {\cal L}}^{(2)}_{\rm Dm}(\dot{\zeta},\zeta,A,B,\dot{\delta\sigma},\delta\sigma)$ is the quadratic Lagrangian of the DHOST part and matter part. The explicit form of the quadratic Lagrangian for the scordatura part ${\tilde {\cal L}}_{\rm S}^{(2)}(\dot{\zeta},\zeta,\dot{A},A,B )$ is given by (\ref{Lagrangian-scordatura}) even after introducing the coupling to the matter. The reason is that the matter is not directly coupled to the scordatura term and also we did not use the background equations in obtaining (\ref{Lagrangian-scordatura}). Thus, our task here is to only find ${\tilde {\cal L}}^{(2)}_{\rm Dm}(\dot{\zeta},\zeta,A,B,\dot{\delta\sigma},\delta\sigma)$.

As we have already fixed the gauge, any scalar mode is gauge-invariant. We thus can work with any combination of these scalar modes. There are many variables for the matter perturbations and we prefer to work with
\begin{align}\label{delta-def}
\delta_{\rm m} &\equiv 
\frac{1}{1+c_{\rm m}^2} \frac{\delta\rho_{\rm m}}{\rho_{\rm m}} + 3 (1+w_{\rm m}) \psi \notag\\
&= (1+w_{\rm m}) \bigg( 3 \zeta - (1+3 \alpha_H c_{\rm m}^2) \frac{A}{c_{\rm m}^2} 
+ \frac{\dot{\delta\sigma}}{c_{\rm m}^2 \dot{\bar\sigma}} \bigg) \,,
\end{align}
where 
\begin{equation}\label{w}
w_{\rm m} \equiv \frac{p_{\rm m}}{\rho_{\rm m}} 
= \frac{{\mathrm p}}{2 {\mathrm y} {\mathrm p}_{,{\mathrm y}} - {\mathrm p}} \,,
\end{equation}
is the equation of state parameter. In obtaining \eqref{delta-def}, we have substituted (\ref{xi-def}) and also we have used the explicit expression of $\delta\rho_{\rm m}$ in the last step. The reason that we prefer to work with the variable \eqref{delta-def} is that it makes the calculations significantly simple.

By the direct calculations, it is straightforward to obtain the quadratic Lagrangian in terms of the k-essence field perturbations ${\tilde {\cal L}}^{(2)}_{\rm Dm}(\dot{\zeta},\zeta,A,B,\dot{\delta\sigma},\delta\sigma)$, while we do not write its explicit form here. We, however, need to find it in terms of the matter density contrast as ${\tilde {\cal L}}^{(2)}_{\rm Dm}(\dot{\zeta},\zeta,A,B,\dot{\delta}_{\rm m},\delta_{\rm m})$ through the transformation (\ref{delta-def}). The transformation (\ref{delta-def}) includes time derivative of the k-essence field $\delta\sigma$ and it is not a point transformation. To perform this transformation we can go to the Hamiltonian formalism and perform the corresponding canonical transformation \cite{Firouzjahi:2017txv}. Equivalently, we can introduce an auxiliary field and then integrate out a non-dynamical field to perform this transformation at the level of Lagrangian \cite{DeFelice:2015moy}. Doing so, we find
\begin{align}\label{Lagrangian-DHOST-m-a}
{\tilde {\cal L}}_{\rm Dm}^{(2)}&=
{\tilde {\cal L}}_{\rm D}^{(2)} + \frac{a^5}{2{\tilde k}^2}\frac{\rho_{\rm m}}{1+w_{\rm m}} 
\Big( \dot{\delta}_{\rm m}^2 
- \frac{c_{\rm m}^2 {\tilde k}^2}{a^2} \delta_{\rm m}^2 
- 3 (c_{\rm m}^2-w_{\rm m}) (\dot{h}+5 h^2 ) \delta_{\rm m}^2 \Big) \nonumber \\ 
&~~~ + a^3 \rho_{\rm m} 
\Big( B \big( \dot{\delta}_{\rm m} + \beta _{\rm v} (c_{\rm m}^2-w_{\rm m}) \delta_{\rm m} \big) 
- ( 1 + 3 \alpha_H c_{\rm m}^2) A \delta_{\rm m} + 3 c_{\rm m}^2 \zeta \delta_{\rm m} \Big) \nonumber \\ 
&~~~ - \frac{1}{2} a^3 \rho_{\rm m} (1+w_{\rm m}) 
\Big( 3 A^2 \beta_{\rm m} - 6 ( 1 + 3 \alpha_H c_{\rm m}^2 ) A \zeta
+ 9 c_{\rm m}^2 \zeta^2 - \frac{{\tilde k}^2}{a^2} B^2 \Big) \,,
\end{align}
where ${\tilde {\cal L}}_{\rm D}^{(2)}$ is given by Eq. \eqref{Lagrangian-DHOST}. Here, we have also defined second order dimensionless parameters 
\begin{align}
\label{beta-m} 
\beta_{\rm m} &\equiv \frac{2 \beta_H 
- 5 \alpha_H^2 - \alpha_H }{1-3 \alpha _H}  + 3 c_{\rm m}^2 \alpha_H^2  \,,  \\
\label{beta-v-f0} 
\beta_{\rm v} &\equiv 3h_b - \frac{3h_b\alpha_H}{1 - 3 \alpha_H }
\frac{ 3 (1+\alpha_B) \alpha_K + \epsilon_{\rm m} ( \alpha_B + 3 \alpha_H)}{
(1+\alpha_B){}^2-\beta_K- \epsilon_{\rm m} \beta_{\rm m}} 
+ \frac{ 9  \epsilon_{\rm m} h_b c_{\rm m}^2\alpha_H^2}{
(1+\alpha_B){}^2-\beta_K- \epsilon_{\rm m} \beta_{\rm m}} \,,
\end{align}
with 
\begin{equation}
\epsilon_{\rm m} \equiv \frac{\rho_{\rm m} (1+w_{\rm m})}{4 f_2 h_b^2} \,,
\end{equation}
which reduces to $\epsilon_{\rm m} = - \frac{\dot{h}}{h^2}$ for the stealth solution. For the case of dark matter with $w_{\rm m} = 0=c_{\rm m}^2$, the quadratic Lagrangian \eqref{Lagrangian-DHOST-m-a} reduces to \eqref{Lagrangian-DHOST-m} which we use to study the linear perturbations for the system of scordatura dark energy coupled to the dark matter.


\bibliographystyle{JHEPmod}
\bibliography{refs}

\providecommand{\href}[2]{#2}\begingroup\raggedright\begin{thebibliography}{10}

\bibitem{Ostrogradsky:1850fid}
M.~Ostrogradsky, {\it {Memoires sur les equations differentielles, relatives au
  probleme des isoperimetres}}, {Mem. Acad. St. Petersbourg {\bfseries 6}
  (1850) 385}.

\bibitem{Woodard:2015zca}
R.~P. Woodard, {\it {Ostrogradsky's theorem on Hamiltonian instability}},
  \href{https://doi.org/10.4249/scholarpedia.32243}{Scholarpedia {\bfseries 10}
  (2015) 32243} [\href{http://arxiv.org/abs/1506.02210}{{\ttfamily
  arXiv:1506.02210}}].

\bibitem{Motohashi:2014opa}
H.~Motohashi and T.~Suyama, {\it {Third order equations of motion and the
  Ostrogradsky instability}},
  \href{https://doi.org/10.1103/PhysRevD.91.085009}{Phys. Rev. {\bfseries D91}
  (2015) 085009} [\href{http://arxiv.org/abs/1411.3721}{{\ttfamily
  arXiv:1411.3721}}].

\bibitem{Motohashi:2020psc}
H.~Motohashi and T.~Suyama, {\it {Quantum Ostrogradsky theorem}},
  \href{https://doi.org/10.1007/JHEP09(2020)032}{JHEP {\bfseries 20} (2020)
  032} [\href{http://arxiv.org/abs/2001.02483}{{\ttfamily arXiv:2001.02483}}].

\bibitem{Aoki:2020gfv}
K.~Aoki and H.~Motohashi, {\it {Ghost from constraints: a generalization of
  Ostrogradsky theorem}},
  \href{https://doi.org/10.1088/1475-7516/2020/08/026}{JCAP {\bfseries 2008}
  (2020) 026} [\href{http://arxiv.org/abs/2001.06756}{{\ttfamily
  arXiv:2001.06756}}].

\bibitem{Langlois:2015cwa}
D.~Langlois and K.~Noui, {\it {Degenerate higher derivative theories beyond
  Horndeski: evading the Ostrogradski instability}},
  \href{https://doi.org/10.1088/1475-7516/2016/02/034}{JCAP {\bfseries 1602}
  (2016) 034} [\href{http://arxiv.org/abs/1510.06930}{{\ttfamily
  arXiv:1510.06930}}].

\bibitem{Motohashi:2016ftl}
H.~Motohashi, K.~Noui, T.~Suyama, M.~Yamaguchi and D.~Langlois, {\it {Healthy
  degenerate theories with higher derivatives}},
  \href{https://doi.org/10.1088/1475-7516/2016/07/033}{JCAP {\bfseries 1607}
  (2016) 033} [\href{http://arxiv.org/abs/1603.09355}{{\ttfamily
  arXiv:1603.09355}}].

\bibitem{Motohashi:2017eya}
H.~Motohashi, T.~Suyama and M.~Yamaguchi, {\it {Ghost-free theory with
  third-order time derivatives}},
  \href{https://doi.org/10.7566/JPSJ.87.063401}{J. Phys. Soc. Jap. {\bfseries
  87} (2018) 063401} [\href{http://arxiv.org/abs/1711.08125}{{\ttfamily
  arXiv:1711.08125}}].

\bibitem{Motohashi:2018pxg}
H.~Motohashi, T.~Suyama and M.~Yamaguchi, {\it {Ghost-free theories with
  arbitrary higher-order time derivatives}},
  \href{https://doi.org/10.1007/JHEP06(2018)133}{JHEP {\bfseries 06} (2018)
  133} [\href{http://arxiv.org/abs/1804.07990}{{\ttfamily arXiv:1804.07990}}].

\bibitem{Crisostomi:2016czh}
M.~Crisostomi, K.~Koyama and G.~Tasinato, {\it {Extended Scalar-Tensor Theories
  of Gravity}}, \href{https://doi.org/10.1088/1475-7516/2016/04/044}{JCAP
  {\bfseries 1604} (2016) 044}
  [\href{http://arxiv.org/abs/1602.03119}{{\ttfamily arXiv:1602.03119}}].

\bibitem{BenAchour:2016fzp}
J.~Ben~Achour, M.~Crisostomi, K.~Koyama, D.~Langlois, K.~Noui and G.~Tasinato,
  {\it {Degenerate higher order scalar-tensor theories beyond Horndeski up to
  cubic order}}, \href{https://doi.org/10.1007/JHEP12(2016)100}{JHEP {\bfseries
  12} (2016) 100} [\href{http://arxiv.org/abs/1608.08135}{{\ttfamily
  arXiv:1608.08135}}].

\bibitem{Horndeski:1974wa}
G.~W. Horndeski, {\it {Second-order scalar-tensor field equations in a
  four-dimensional space}}, \href{https://doi.org/10.1007/BF01807638}{Int. J.
  Theor. Phys. {\bfseries 10} (1974) 363}.

\bibitem{Zumalacarregui:2013pma}
M.~Zumalac\'arregui and J.~Garc\'ia-Bellido, {\it {Transforming gravity: from
  derivative couplings to matter to second-order scalar-tensor theories beyond
  the Horndeski Lagrangian}},
  \href{https://doi.org/10.1103/PhysRevD.89.064046}{Phys. Rev. {\bfseries D89}
  (2014) 064046} [\href{http://arxiv.org/abs/1308.4685}{{\ttfamily
  arXiv:1308.4685}}].

\bibitem{Gleyzes:2014dya}
J.~Gleyzes, D.~Langlois, F.~Piazza and F.~Vernizzi, {\it {Healthy theories
  beyond Horndeski}},
  \href{https://doi.org/10.1103/PhysRevLett.114.211101}{Phys. Rev. Lett.
  {\bfseries 114} (2015) 211101}
  [\href{http://arxiv.org/abs/1404.6495}{{\ttfamily arXiv:1404.6495}}].

\bibitem{Gao:2014soa}
X.~Gao, {\it {Unifying framework for scalar-tensor theories of gravity}},
  \href{https://doi.org/10.1103/PhysRevD.90.081501}{Phys. Rev. {\bfseries D90}
  (2014) 081501} [\href{http://arxiv.org/abs/1406.0822}{{\ttfamily
  arXiv:1406.0822}}].

\bibitem{DeFelice:2018mkq}
A.~De~Felice, D.~Langlois, S.~Mukohyama, K.~Noui and A.~Wang, {\it {Generalized
  instantaneous modes in higher-order scalar-tensor theories}},
  \href{https://doi.org/10.1103/PhysRevD.98.084024}{Phys. Rev. {\bfseries D98}
  (2018) 084024} [\href{http://arxiv.org/abs/1803.06241}{{\ttfamily
  arXiv:1803.06241}}].

\bibitem{Gao:2018znj}
X.~Gao and Z.-B. Yao, {\it {Spatially covariant gravity with velocity of the
  lapse function: the Hamiltonian analysis}},
  \href{https://doi.org/10.1088/1475-7516/2019/05/024}{JCAP {\bfseries 1905}
  (2019) 024} [\href{http://arxiv.org/abs/1806.02811}{{\ttfamily
  arXiv:1806.02811}}].

\bibitem{Motohashi:2020wxj}
H.~Motohashi and W.~Hu, {\it {Effective field theory of degenerate higher-order
  inflation}}, \href{https://doi.org/10.1103/PhysRevD.101.083531}{Phys. Rev. D
  {\bfseries 101} (2020) 083531}
  [\href{http://arxiv.org/abs/2002.07967}{{\ttfamily arXiv:2002.07967}}].

\bibitem{Zumalacarregui:2012us}
M.~Zumalacarregui, T.~S. Koivisto and D.~F. Mota, {\it {DBI Galileons in the
  Einstein Frame: Local Gravity and Cosmology}},
  \href{https://doi.org/10.1103/PhysRevD.87.083010}{Phys. Rev. D {\bfseries 87}
  (2013) 083010} [\href{http://arxiv.org/abs/1210.8016}{{\ttfamily
  arXiv:1210.8016}}].

\bibitem{Bettoni:2013diz}
D.~Bettoni and S.~Liberati, {\it {Disformal invariance of second order
  scalar-tensor theories: Framing the Horndeski action}},
  \href{https://doi.org/10.1103/PhysRevD.88.084020}{Phys. Rev. D {\bfseries 88}
  (2013) 084020} [\href{http://arxiv.org/abs/1306.6724}{{\ttfamily
  arXiv:1306.6724}}].

\bibitem{Achour:2016rkg}
J.~Ben~Achour, D.~Langlois and K.~Noui, {\it {Degenerate higher order
  scalar-tensor theories beyond Horndeski and disformal transformations}},
  \href{https://doi.org/10.1103/PhysRevD.93.124005}{Phys. Rev. {\bfseries D93}
  (2016) 124005} [\href{http://arxiv.org/abs/1602.08398}{{\ttfamily
  arXiv:1602.08398}}].

\bibitem{Domenech:2015tca}
G.~Dom\`enech, S.~Mukohyama, R.~Namba, A.~Naruko, R.~Saitou and Y.~Watanabe,
  {\it {Derivative-dependent metric transformation and physical degrees of
  freedom}}, \href{https://doi.org/10.1103/PhysRevD.92.084027}{Phys. Rev.
  {\bfseries D92} (2015) 084027}
  [\href{http://arxiv.org/abs/1507.05390}{{\ttfamily arXiv:1507.05390}}].

\bibitem{Takahashi:2017zgr}
K.~Takahashi, H.~Motohashi, T.~Suyama and T.~Kobayashi, {\it {General
  invertible transformation and physical degrees of freedom}},
  \href{https://doi.org/10.1103/PhysRevD.95.084053}{Phys. Rev. {\bfseries D95}
  (2017) 084053} [\href{http://arxiv.org/abs/1702.01849}{{\ttfamily
  arXiv:1702.01849}}].

\bibitem{Deruelle:2014zza}
N.~Deruelle and J.~Rua, {\it {Disformal Transformations, Veiled General
  Relativity and Mimetic Gravity}},
  \href{https://doi.org/10.1088/1475-7516/2014/09/002}{JCAP {\bfseries 09}
  (2014) 002} [\href{http://arxiv.org/abs/1407.0825}{{\ttfamily
  arXiv:1407.0825}}].

\bibitem{Chamseddine:2013kea}
A.~H. Chamseddine and V.~Mukhanov, {\it {Mimetic Dark Matter}},
  \href{https://doi.org/10.1007/JHEP11(2013)135}{JHEP {\bfseries 11} (2013)
  135} [\href{http://arxiv.org/abs/1308.5410}{{\ttfamily arXiv:1308.5410}}].

\bibitem{Takahashi:2017pje}
K.~Takahashi and T.~Kobayashi, {\it {Extended mimetic gravity: Hamiltonian
  analysis and gradient instabilities}},
  \href{https://doi.org/10.1088/1475-7516/2017/11/038}{JCAP {\bfseries 1711}
  (2017) 038} [\href{http://arxiv.org/abs/1708.02951}{{\ttfamily
  arXiv:1708.02951}}].

\bibitem{Langlois:2018jdg}
D.~Langlois, M.~Mancarella, K.~Noui and F.~Vernizzi, {\it {Mimetic gravity as
  DHOST theories}}, \href{https://doi.org/10.1088/1475-7516/2019/02/036}{JCAP
  {\bfseries 1902} (2019) 036}
  [\href{http://arxiv.org/abs/1802.03394}{{\ttfamily arXiv:1802.03394}}].

\bibitem{Firouzjahi:2017txv}
H.~Firouzjahi, M.~A. Gorji and S.~A. Hosseini~Mansoori, {\it {Instabilities in
  Mimetic Matter Perturbations}},
  \href{https://doi.org/10.1088/1475-7516/2017/07/031}{JCAP {\bfseries 07}
  (2017) 031} [\href{http://arxiv.org/abs/1703.02923}{{\ttfamily
  arXiv:1703.02923}}].

\bibitem{Zheng:2017qfs}
Y.~Zheng, L.~Shen, Y.~Mou and M.~Li, {\it {On (in)stabilities of perturbations
  in mimetic models with higher derivatives}},
  \href{https://doi.org/10.1088/1475-7516/2017/08/040}{JCAP {\bfseries 08}
  (2017) 040} [\href{http://arxiv.org/abs/1704.06834}{{\ttfamily
  arXiv:1704.06834}}].

\bibitem{Hirano:2017zox}
S.~Hirano, S.~Nishi and T.~Kobayashi, {\it {Healthy imperfect dark matter from
  effective theory of mimetic cosmological perturbations}},
  \href{https://doi.org/10.1088/1475-7516/2017/07/009}{JCAP {\bfseries 07}
  (2017) 009} [\href{http://arxiv.org/abs/1704.06031}{{\ttfamily
  arXiv:1704.06031}}].

\bibitem{Gorji:2017cai}
M.~A. Gorji, S.~A. Hosseini~Mansoori and H.~Firouzjahi, {\it {Higher Derivative
  Mimetic Gravity}}, \href{https://doi.org/10.1088/1475-7516/2018/01/020}{JCAP
  {\bfseries 01} (2018) 020} [\href{http://arxiv.org/abs/1709.09988}{{\ttfamily
  arXiv:1709.09988}}].

\bibitem{Mukohyama:2009tp}
S.~Mukohyama, {\it {Caustic avoidance in Horava-Lifshitz gravity}},
  \href{https://doi.org/10.1088/1475-7516/2009/09/005}{JCAP {\bfseries 09}
  (2009) 005} [\href{http://arxiv.org/abs/0906.5069}{{\ttfamily
  arXiv:0906.5069}}].

\bibitem{Barvinsky:2013mea}
A.~Barvinsky, {\it {Dark matter as a ghost free conformal extension of Einstein
  theory}}, \href{https://doi.org/10.1088/1475-7516/2014/01/014}{JCAP
  {\bfseries 01} (2014) 014} [\href{http://arxiv.org/abs/1311.3111}{{\ttfamily
  arXiv:1311.3111}}].

\bibitem{Gorji:2019rlm}
M.~A. Gorji, A.~Allahyari, M.~Khodadi and H.~Firouzjahi, {\it {Mimetic black
  holes}}, \href{https://doi.org/10.1103/PhysRevD.101.124060}{Phys. Rev. D
  {\bfseries 101} (2020) 124060}
  [\href{http://arxiv.org/abs/1912.04636}{{\ttfamily arXiv:1912.04636}}].

\bibitem{Motohashi:2018wdq}
H.~Motohashi and M.~Minamitsuji, {\it {General Relativity solutions in modified
  gravity}}, \href{https://doi.org/10.1016/j.physletb.2018.04.041}{Phys. Lett.
  {\bfseries B781} (2018) 728}
  [\href{http://arxiv.org/abs/1804.01731}{{\ttfamily arXiv:1804.01731}}].

\bibitem{Takahashi:2020hso}
K.~Takahashi and H.~Motohashi, {\it {General Relativity solutions with stealth
  scalar hair in quadratic higher-order scalar-tensor theories}},
  \href{https://doi.org/10.1088/1475-7516/2020/06/034}{JCAP {\bfseries 06}
  (2020) 034} [\href{http://arxiv.org/abs/2004.03883}{{\ttfamily
  arXiv:2004.03883}}].

\bibitem{ArkaniHamed:2003uy}
N.~Arkani-Hamed, H.-C. Cheng, M.~A. Luty and S.~Mukohyama, {\it {Ghost
  condensation and a consistent infrared modification of gravity}},
  \href{https://doi.org/10.1088/1126-6708/2004/05/074}{JHEP {\bfseries 05}
  (2004) 074} [\href{http://arxiv.org/abs/hep-th/0312099}{{\ttfamily
  arXiv:hep-th/0312099}}].

\bibitem{Mukohyama:2005rw}
S.~Mukohyama, {\it {Black holes in the ghost condensate}},
  \href{https://doi.org/10.1103/PhysRevD.71.104019}{Phys. Rev. {\bfseries D71}
  (2005) 104019} [\href{http://arxiv.org/abs/hep-th/0502189}{{\ttfamily
  arXiv:hep-th/0502189}}].

\bibitem{Crisostomi:2018bsp}
M.~Crisostomi, K.~Koyama, D.~Langlois, K.~Noui and D.~Steer, {\it {Cosmological
  evolution in DHOST theories}},
  \href{https://doi.org/10.1088/1475-7516/2019/01/030}{JCAP {\bfseries 01}
  (2019) 030} [\href{http://arxiv.org/abs/1810.12070}{{\ttfamily
  arXiv:1810.12070}}].

\bibitem{Babichev:2013cya}
E.~Babichev and C.~Charmousis, {\it {Dressing a black hole with a
  time-dependent Galileon}},
  \href{https://doi.org/10.1007/JHEP08(2014)106}{JHEP {\bfseries 08} (2014)
  106} [\href{http://arxiv.org/abs/1312.3204}{{\ttfamily arXiv:1312.3204}}].

\bibitem{Minamitsuji:2018vuw}
M.~Minamitsuji and H.~Motohashi, {\it {Stealth Schwarzschild solution in shift
  symmetry breaking theories}},
  \href{https://doi.org/10.1103/PhysRevD.98.084027}{Phys. Rev. {\bfseries D98}
  (2018) 084027} [\href{http://arxiv.org/abs/1809.06611}{{\ttfamily
  arXiv:1809.06611}}].

\bibitem{BenAchour:2018dap}
J.~Ben~Achour and H.~Liu, {\it {Hairy Schwarzschild-(A)dS black hole solutions
  in DHOST theories beyond shift symmetry}},
  \href{https://doi.org/10.1103/PhysRevD.99.064042}{Phys. Rev. {\bfseries D99}
  (2019) 064042} [\href{http://arxiv.org/abs/1811.05369}{{\ttfamily
  arXiv:1811.05369}}].

\bibitem{Motohashi:2019sen}
H.~Motohashi and M.~Minamitsuji, {\it {Exact black hole solutions in
  shift-symmetric quadratic degenerate higher-order scalar-tensor theories}},
  \href{https://doi.org/10.1103/PhysRevD.99.064040}{Phys. Rev. {\bfseries D99}
  (2019) 064040} [\href{http://arxiv.org/abs/1901.04658}{{\ttfamily
  arXiv:1901.04658}}].

\bibitem{Charmousis:2019vnf}
C.~Charmousis, M.~Crisostomi, R.~Gregory and N.~Stergioulas, {\it {Rotating
  Black Holes in Higher Order Gravity}},
  \href{https://doi.org/10.1103/PhysRevD.100.084020}{Phys. Rev. {\bfseries
  D100} (2019) 084020} [\href{http://arxiv.org/abs/1903.05519}{{\ttfamily
  arXiv:1903.05519}}].

\bibitem{Minamitsuji:2019shy}
M.~Minamitsuji and J.~Edholm, {\it {Black hole solutions in shift-symmetric
  degenerate higher-order scalar-tensor theories}},
  \href{https://doi.org/10.1103/PhysRevD.100.044053}{Phys. Rev. {\bfseries
  D100} (2019) 044053} [\href{http://arxiv.org/abs/1907.02072}{{\ttfamily
  arXiv:1907.02072}}].

\bibitem{BenAchour:2019fdf}
J.~Ben~Achour, H.~Liu and S.~Mukohyama, {\it {Hairy black holes in DHOST
  theories: Exploring disformal transformation as a solution-generating
  method}}, \href{https://doi.org/10.1088/1475-7516/2020/02/023}{JCAP
  {\bfseries 2002} (2020) 023}
  [\href{http://arxiv.org/abs/1910.11017}{{\ttfamily arXiv:1910.11017}}].

\bibitem{Minamitsuji:2019tet}
M.~Minamitsuji and J.~Edholm, {\it {Black holes with a nonconstant kinetic term
  in degenerate higher-order scalar tensor theories}},
  \href{https://doi.org/10.1103/PhysRevD.101.044034}{Phys. Rev. {\bfseries
  D101} (2020) 044034} [\href{http://arxiv.org/abs/1912.01744}{{\ttfamily
  arXiv:1912.01744}}].

\bibitem{Bernardo:2019yxp}
R.~C. Bernardo, J.~Celestial and I.~Vega, {\it {Stealth black holes in shift
  symmetric kinetic gravity braiding}},
  \href{https://doi.org/10.1103/PhysRevD.101.024036}{Phys. Rev. D {\bfseries
  101} (2020) 024036} [\href{http://arxiv.org/abs/1911.01847}{{\ttfamily
  arXiv:1911.01847}}].

\bibitem{Takahashi:2019oxz}
K.~Takahashi, H.~Motohashi and M.~Minamitsuji, {\it {Linear stability analysis
  of hairy black holes in quadratic degenerate higher-order scalar-tensor
  theories: Odd-parity perturbations}},
  \href{https://doi.org/10.1103/PhysRevD.100.024041}{Phys. Rev. {\bfseries
  D100} (2019) 024041} [\href{http://arxiv.org/abs/1904.03554}{{\ttfamily
  arXiv:1904.03554}}].

\bibitem{BenAchour:2020fgy}
J.~Ben~Achour, H.~Liu, H.~Motohashi, S.~Mukohyama and K.~Noui, {\it {On
  rotating black holes in DHOST theories}},
  \href{https://doi.org/10.1088/1475-7516/2020/11/001}{JCAP {\bfseries 11}
  (2020) 001} [\href{http://arxiv.org/abs/2006.07245}{{\ttfamily
  arXiv:2006.07245}}].

\bibitem{deRham:2019slh}
C.~de~Rham and J.~Zhang, {\it {Perturbations of stealth black holes in
  degenerate higher-order scalar-tensor theories}},
  \href{https://doi.org/10.1103/PhysRevD.100.124023}{Phys. Rev. {\bfseries
  D100} (2019) 124023} [\href{http://arxiv.org/abs/1907.00699}{{\ttfamily
  arXiv:1907.00699}}].

\bibitem{Motohashi:2019ymr}
H.~Motohashi and S.~Mukohyama, {\it {Weakly-coupled stealth solution in
  scordatura degenerate theory}},
  \href{https://doi.org/10.1088/1475-7516/2020/01/030}{JCAP {\bfseries 2001}
  (2020) 030} [\href{http://arxiv.org/abs/1912.00378}{{\ttfamily
  arXiv:1912.00378}}].

\bibitem{Khoury:2020aya}
J.~Khoury, M.~Trodden and S.~S. Wong, {\it {Existence and Instability of Novel
  Hairy Black Holes in Shift-symmetric Horndeski Theories}},
  \href{https://doi.org/10.1088/1475-7516/2020/11/044}{JCAP {\bfseries 11}
  (2020) 044} [\href{http://arxiv.org/abs/2007.01320}{{\ttfamily
  arXiv:2007.01320}}].

\bibitem{Langlois:2015skt}
D.~Langlois and K.~Noui, {\it {Hamiltonian analysis of higher derivative
  scalar-tensor theories}},
  \href{https://doi.org/10.1088/1475-7516/2016/07/016}{JCAP {\bfseries 07}
  (2016) 016} [\href{http://arxiv.org/abs/1512.06820}{{\ttfamily
  arXiv:1512.06820}}].

\bibitem{Langlois:2017mxy}
D.~Langlois, M.~Mancarella, K.~Noui and F.~Vernizzi, {\it {Effective
  Description of Higher-Order Scalar-Tensor Theories}},
  \href{https://doi.org/10.1088/1475-7516/2017/05/033}{JCAP {\bfseries 1705}
  (2017) 033} [\href{http://arxiv.org/abs/1703.03797}{{\ttfamily
  arXiv:1703.03797}}].

\bibitem{Horava:2009uw}
P.~Horava, {\it {Quantum Gravity at a Lifshitz Point}},
  \href{https://doi.org/10.1103/PhysRevD.79.084008}{Phys. Rev. {\bfseries D79}
  (2009) 084008} [\href{http://arxiv.org/abs/0901.3775}{{\ttfamily
  arXiv:0901.3775}}].

\bibitem{Mukohyama:2009rk}
S.~Mukohyama, {\it {Ghost condensate and generalized second law}},
  \href{https://doi.org/10.1088/1126-6708/2009/09/070}{JHEP {\bfseries 09}
  (2009) 070} [\href{http://arxiv.org/abs/0901.3595}{{\ttfamily
  arXiv:0901.3595}}].

\bibitem{Mukohyama:2009um}
S.~Mukohyama, {\it {Can ghost condensate decrease entropy?}},
  \href{https://doi.org/10.2174/1874381101003020030,
  10.2174/1874381101003010030}{Open Astron. J. {\bfseries 3} (2010) 30}
  [\href{http://arxiv.org/abs/0908.4123}{{\ttfamily arXiv:0908.4123}}].

\bibitem{Mukohyama:2006be}
S.~Mukohyama, {\it {Accelerating Universe and Cosmological Perturbation in the
  Ghost Condensate}}, \href{https://doi.org/10.1088/1475-7516/2006/10/011}{JCAP
  {\bfseries 0610} (2006) 011}
  [\href{http://arxiv.org/abs/hep-th/0607181}{{\ttfamily
  arXiv:hep-th/0607181}}].

\bibitem{ArmendarizPicon:1999rj}
C.~Armendariz-Picon, T.~Damour and V.~F. Mukhanov, {\it {k - inflation}},
  \href{https://doi.org/10.1016/S0370-2693(99)00603-6}{Phys. Lett. B {\bfseries
  458} (1999) 209} [\href{http://arxiv.org/abs/hep-th/9904075}{{\ttfamily
  arXiv:hep-th/9904075}}].

\bibitem{Deffayet:2010qz}
C.~Deffayet, O.~Pujolas, I.~Sawicki and A.~Vikman, {\it {Imperfect Dark Energy
  from Kinetic Gravity Braiding}},
  \href{https://doi.org/10.1088/1475-7516/2010/10/026}{JCAP {\bfseries 10}
  (2010) 026} [\href{http://arxiv.org/abs/1008.0048}{{\ttfamily
  arXiv:1008.0048}}].

\bibitem{Langlois:2017dyl}
D.~Langlois, R.~Saito, D.~Yamauchi and K.~Noui, {\it {Scalar-tensor theories
  and modified gravity in the wake of GW170817}},
  \href{https://doi.org/10.1103/PhysRevD.97.061501}{Phys. Rev. {\bfseries D97}
  (2018) 061501} [\href{http://arxiv.org/abs/1711.07403}{{\ttfamily
  arXiv:1711.07403}}].

\bibitem{Monitor:2017mdv}
{\scshape LIGO Scientific, Virgo, Fermi-GBM, INTEGRAL} collaboration, B.~P.
  Abbott et~al., {\it {Gravitational Waves and Gamma-rays from a Binary Neutron
  Star Merger: GW170817 and GRB 170817A}},
  \href{https://doi.org/10.3847/2041-8213/aa920c}{Astrophys. J. {\bfseries 848}
  (2017) L13} [\href{http://arxiv.org/abs/1710.05834}{{\ttfamily
  arXiv:1710.05834}}].

\bibitem{Creminelli:2018xsv}
P.~Creminelli, M.~Lewandowski, G.~Tambalo and F.~Vernizzi, {\it {Gravitational
  Wave Decay into Dark Energy}},
  \href{https://doi.org/10.1088/1475-7516/2018/12/025}{JCAP {\bfseries 1812}
  (2018) 025} [\href{http://arxiv.org/abs/1809.03484}{{\ttfamily
  arXiv:1809.03484}}].

\bibitem{Bellini:2014fua}
E.~Bellini and I.~Sawicki, {\it {Maximal freedom at minimum cost: linear
  large-scale structure in general modifications of gravity}},
  \href{https://doi.org/10.1088/1475-7516/2014/07/050}{JCAP {\bfseries 07}
  (2014) 050} [\href{http://arxiv.org/abs/1404.3713}{{\ttfamily
  arXiv:1404.3713}}].

\bibitem{Gleyzes:2014rba}
J.~Gleyzes, D.~Langlois and F.~Vernizzi, {\it {A unifying description of dark
  energy}}, \href{https://doi.org/10.1142/S021827181443010X}{Int. J. Mod. Phys.
  D {\bfseries 23} (2015) 1443010}
  [\href{http://arxiv.org/abs/1411.3712}{{\ttfamily arXiv:1411.3712}}].

\bibitem{Motohashi:2017gqb}
H.~Motohashi and W.~Hu, {\it {Generalized Slow Roll in the Unified Effective
  Field Theory of Inflation}},
  \href{https://doi.org/10.1103/PhysRevD.96.023502}{Phys. Rev. D {\bfseries 96}
  (2017) 023502} [\href{http://arxiv.org/abs/1704.01128}{{\ttfamily
  arXiv:1704.01128}}].

\bibitem{ArkaniHamed:2005gu}
N.~Arkani-Hamed, H.-C. Cheng, M.~A. Luty, S.~Mukohyama and T.~Wiseman, {\it
  {Dynamics of gravity in a Higgs phase}},
  \href{https://doi.org/10.1088/1126-6708/2007/01/036}{JHEP {\bfseries 01}
  (2007) 036} [\href{http://arxiv.org/abs/hep-ph/0507120}{{\ttfamily
  arXiv:hep-ph/0507120}}].

\bibitem{Gumrukcuoglu:2016jbh}
A.~E. Gümrükçüo\u{g}lu, S.~Mukohyama and T.~P. Sotiriou, {\it {Low energy
  ghosts and the Jeans' instability}},
  \href{https://doi.org/10.1103/PhysRevD.94.064001}{Phys. Rev. D {\bfseries 94}
  (2016) 064001} [\href{http://arxiv.org/abs/1606.00618}{{\ttfamily
  arXiv:1606.00618}}].

\bibitem{Sawicki:2015zya}
I.~Sawicki and E.~Bellini, {\it {Limits of quasistatic approximation in
  modified-gravity cosmologies}},
  \href{https://doi.org/10.1103/PhysRevD.92.084061}{Phys. Rev. D {\bfseries 92}
  (2015) 084061} [\href{http://arxiv.org/abs/1503.06831}{{\ttfamily
  arXiv:1503.06831}}].

\bibitem{Gleyzes:2015pma}
J.~Gleyzes, D.~Langlois, M.~Mancarella and F.~Vernizzi, {\it {Effective Theory
  of Interacting Dark Energy}},
  \href{https://doi.org/10.1088/1475-7516/2015/08/054}{JCAP {\bfseries 08}
  (2015) 054} [\href{http://arxiv.org/abs/1504.05481}{{\ttfamily
  arXiv:1504.05481}}].

\bibitem{Tsujikawa:2007gd}
S.~Tsujikawa, {\it {Matter density perturbations and effective gravitational
  constant in modified gravity models of dark energy}},
  \href{https://doi.org/10.1103/PhysRevD.76.023514}{Phys. Rev. D {\bfseries 76}
  (2007) 023514} [\href{http://arxiv.org/abs/0705.1032}{{\ttfamily
  arXiv:0705.1032}}].

\bibitem{DeFelice:2010as}
A.~De~Felice, R.~Kase and S.~Tsujikawa, {\it {Matter perturbations in Galileon
  cosmology}}, \href{https://doi.org/10.1103/PhysRevD.83.043515}{Phys. Rev. D
  {\bfseries 83} (2011) 043515}
  [\href{http://arxiv.org/abs/1011.6132}{{\ttfamily arXiv:1011.6132}}].

\bibitem{DeFelice:2011hq}
A.~De~Felice, T.~Kobayashi and S.~Tsujikawa, {\it {Effective gravitational
  couplings for cosmological perturbations in the most general scalar-tensor
  theories with second-order field equations}},
  \href{https://doi.org/10.1016/j.physletb.2011.11.028}{Phys. Lett. B
  {\bfseries 706} (2011) 123} [\href{http://arxiv.org/abs/1108.4242}{{\ttfamily
  arXiv:1108.4242}}].

\bibitem{DeFelice:2011aa}
A.~De~Felice and S.~Tsujikawa, {\it {Cosmological constraints on extended
  Galileon models}}, \href{https://doi.org/10.1088/1475-7516/2012/03/025}{JCAP
  {\bfseries 03} (2012) 025} [\href{http://arxiv.org/abs/1112.1774}{{\ttfamily
  arXiv:1112.1774}}].

\bibitem{DAmico:2016ntq}
G.~D'Amico, Z.~Huang, M.~Mancarella and F.~Vernizzi, {\it {Weakening Gravity on
  Redshift-Survey Scales with Kinetic Matter Mixing}},
  \href{https://doi.org/10.1088/1475-7516/2017/02/014}{JCAP {\bfseries 02}
  (2017) 014} [\href{http://arxiv.org/abs/1609.01272}{{\ttfamily
  arXiv:1609.01272}}].

\bibitem{Kase:2020hst}
R.~Kase and S.~Tsujikawa, {\it {General formulation of cosmological
  perturbations in scalar-tensor dark energy coupled to dark matter}},
  \href{https://doi.org/10.1088/1475-7516/2020/11/032}{JCAP {\bfseries 11}
  (2020) 032} [\href{http://arxiv.org/abs/2005.13809}{{\ttfamily
  arXiv:2005.13809}}].

\bibitem{Kobayashi:2014ida}
T.~Kobayashi, Y.~Watanabe and D.~Yamauchi, {\it {Breaking of Vainshtein
  screening in scalar-tensor theories beyond Horndeski}},
  \href{https://doi.org/10.1103/PhysRevD.91.064013}{Phys. Rev. D {\bfseries 91}
  (2015) 064013} [\href{http://arxiv.org/abs/1411.4130}{{\ttfamily
  arXiv:1411.4130}}].

\bibitem{Nutma:2013zea}
T.~Nutma, {\it {xTras : A field-theory inspired xAct package for mathematica}},
  \href{https://doi.org/10.1016/j.cpc.2014.02.006}{Comput. Phys. Commun.
  {\bfseries 185} (2014) 1719}
  [\href{http://arxiv.org/abs/1308.3493}{{\ttfamily arXiv:1308.3493}}].

\bibitem{DeFelice:2015moy}
A.~De~Felice and S.~Mukohyama, {\it {Phenomenology in minimal theory of massive
  gravity}}, \href{https://doi.org/10.1088/1475-7516/2016/04/028}{JCAP
  {\bfseries 04} (2016) 028} [\href{http://arxiv.org/abs/1512.04008}{{\ttfamily
  arXiv:1512.04008}}].

\end{thebibliography}\endgroup

\end{document}